\documentclass[aps,prb,twocolumn,showpacs,10pt]{revtex4-1}
\usepackage{graphicx}
\usepackage{amssymb}
\usepackage{amsmath,color,mathrsfs}
\usepackage{bbold}
\usepackage{comment}
\usepackage[dvipsnames]{xcolor}
\usepackage{bm}
\pdfoutput=1
\usepackage{hyperref}
\hypersetup{urlcolor=blue, colorlinks=true, citecolor=blue, linkcolor=blue}

\usepackage{todonotes}
\usepackage{mdframed}
\graphicspath{{./imgs/}}

\newcommand{\gOne}{g}

\begin{document}
	
\title{Degeneracy lifting of Majorana bound states due to electron-phonon interactions} 

\author{Pavel P. Aseev$^{1}$}
\author{Pasquale Marra$^{2}$}
\author{Peter Stano$^{2,3,4}$}
\author{Jelena Klinovaja$^{1}$}
\author{Daniel Loss$^{1,2}$}

\affiliation{$^{1}$Department of Physics, University of Basel, Klingelbergstrasse 82, CH-4056 Basel, Switzerland}
\affiliation{$^{2}$RIKEN Center for Emergent Matter Science (CEMS), Wako, Saitama 351-0198, Japan}
\affiliation{$^{3}$Department of Applied Physics, School of Engineering, University of Tokyo, 7-3-1 Hongo, Bunkyo-ku, Tokyo 113-8656, Japan}
\affiliation{$^{4}$Institute of Physics, Slovak Academy of Sciences, 845 11 Bratislava, Slovakia}

\date{\today}

\begin{abstract}	
		We study theoretically how electron-phonon interaction affects the energies and level broadening (inverse lifetime) of Majorana bound states (MBSs) in a clean topological nanowire at low temperatures. 
		At zero temperature, the energy splitting between the right and left MBSs remains exponentially small with increasing nanowire length $L$. 
		At finite temperatures, however, the absorption of thermal phonons leads to the broadening of energy levels of the MBSs that does not decay with system length, and the coherent absorption/emission of phonons at  opposite ends of the nanowire results in MBSs energy splitting that decays only as an inverse power-law in $L$. Both effects remain exponential in temperature. In the case of quantized transverse motion of phonons, the presence of Van Hove singularities in the phonon density of states causes additional resonant enhancement of both the energy splitting and the level broadening of the MBSs. This is the most favorable case to observe the phonon-induced energy splitting of MBSs as it becomes much larger than the broadening even if the topological nanowire is much longer than the coherence length.
		We also calculate the charge and spin associated with the energy splitting of the MBSs induced by phonons.
		We consider both a spinless low-energy continuum model, which we evaluate analytically, as well as a spinful lattice model for a Rashba nanowire, which we evaluate numerically.
	\end{abstract}
	
	\maketitle
	
	\section{Introduction}

Topological quantum computing~\cite{Kitaev2001, Kitaev2003, Nayak2008, Wilczek2009, Stern} allows one to encode quantum information in degenerate many-body ground states in such a way that it is virtually free of decoherence. 
Such topologically protected ground states can be implemented as zero-energy Majorana bound states (MBSs) in a topological superconductor (TSC)~\cite{Schnyder2008, Sato2009, Qi2011, Tanaka2012, Vijay2015, Vijay2016}. In one-dimensional (1D) topological superconductors, the MBSs  localized at the opposite ends of the system form a single fermionic state, which
is a highly non-local quantum state robust against local perturbations. The MBS energy is zero (with respect to the chemical potential of the system) in the thermodynamic limit at zero temperature. 
For a finite system, it is exponentially small in the system size\cite{Kitaev2001},
and at a finite temperature, the thermal fluctuations are exponentially suppressed by the superconducting gap\cite{Cheng2012,Bonderson2013,Sarma2015}. 
Indications of the existence of MBSs have been observed in iron atomic chains on the surface of a superconductor\cite{Nadj-Perge2014,Pawlak2016,Ruby2015}. 
One of the most promising systems are semiconducting Rashba nanowires (NWs) in proximity with an
$s$-wave superconductor and in the presence of magnetic fields~\cite{Alicea2010, Lutchyn2010, Oreg2010, Mourik2012,  Das2012, Deng2012, Sticlet2012, Rokhinson2012,  Klinovaja2012, Chevallier2012, San-Jose2012, Dominguez2012, Terhal2012, Klinovaja2012a, Prada2012, Churchill2013,  DeGottardi2013, Thakurathi2013, Maier2014,  Escribano2017, Prada2017, Ptok2017, Kobialka2018, DeMoor2018}. 

However, in a real world environment, MBSs are affected by decoherence as soon as they become coupled with an external system. In this case the TSC may be driven out of its ground state. This happens if manipulations with MBSs are performed non-adiabatically~\cite{Scheurer2013, Sekania2017} or if the TSC is coupled to ungapped~\cite{Budich2012} or gapped~\cite{Goldstein2011, Rainis2012} fermionic baths as well as to fluctuating bosonic fields  \cite{Goldstein2011, Pedrochi2015, Pedrochi2015b} (\textit{e.g.} phonons  \cite{Goldstein2011, Knapp2018}, thermal fluctuations of a gate potential~\cite{Schmidt2012, Lai2018, Aseev2018}, and or electromagnetic environments~\cite{Knapp2018}). In particular, Rainis~\emph{et al.}\cite{Rainis2012} have shown that MBSs have a finite lifetime due to quasiparticle poisoning, i.e., due to a finite tunneling rate of superconducting quasiparticles from the bulk superconductor to the wire.
Budich~\emph{et al.}\cite{Budich2012} have demonstrated that another source of decoherence is the tunneling to a non-gapped fermionic bath, such as a quantum dot or a metallic lead used as a gate or a tip to probe or to perform operations on the wire.
In these two cases the finite lifetime originates from the fact that external coupling explicitly breaks the parity of the fermionic ground state.
However, even if parity is preserved, decoherence can arise from the coupling to a gapped fermionic bath through a bosonic field, in the case that the fermionic and bosonic spectra overlap, as shown by Goldstein~\emph{et al.}\cite{Goldstein2011}
This coupling can originate, for example, if the wire is capacitatively coupled to a metallic or semiconducting gate.
In this case the charge fluctuations at the gates can induce decoherence at finite temperature, as shown by Schmidt~\emph{et al.}\cite{Schmidt2012}
In all these cases, the MBS is coupled with a fermionic bath that is external to the wire.

In this work, we consider the effect of electron-phonon interactions on the MBS, as a fundamental and ubiquitous ingredient of any condensed matter system or nanodevice. Thus, this source of decoherence is intrinsically built-in in the topological wire rather than provided by the external environment or an external physical device. In particular, we calculate the energy splitting between the MBSs and the energy level broadening caused by transitions from MBSs to delocalized bulk states with absorbtion of thermal phonons. The level broadening, expressed as  rate, determines the lifetime of a quasiparticle occupying the levels formed by the two MBSs. We consider also the case when the motion of phonons is quantized in transverse direction to the nanowire (confined phonons), focusing on the situation where the bottom of one of the phonon branches is close to a quasiparticle gap of a 1D TSC. We study how the presence of Van Hove singularities (VHS) \cite{Cleland2002,Trif2008,Kloeffel2014} at the bottom of the phonon spectrum affects the energies of the MBSs. We  consider first a spinless low-energy continuum model  which we can treat largely analytically, and then we also consider a spinful lattice Rashba model numerically where we find very similar results.
	
The outline of the paper is as follows. In Sec.~\ref{sec:model} we introduce a low energy model of a 1D topological nanowire. In Sec.~\ref{sec:1D-acoustic} we discuss the effects of the electron-phonon interaction for 1D acoustic phonons. In Sec.~\ref{sec:quasi-1D} we consider how the interaction with confined phonons with VHS affects the energies of the MBSs, as well as the charge and spin of the perturbed MBSs. Finally we conclude with Sec.~\ref{sec:conclusions}. Technical details are deferred to Apps.~\ref{app:BdG}{}-\ref{app:estimations}.

	\section{The model\label{sec:model}}
	We consider a clean topological nanowire of length $L$\label{def:L} aligned along the $x$-axis\label{def:x} and coupled to a phonon bath.
In order to model the electron subsystem we consider 1D spinless electrons described by the field operator $\Psi(x)$\label{def:Psi}. We define the fermionic fields $\Psi_{\eta}$ \label{def:Psi_eta} with $\eta=+$ ($\eta=-$) corresponding to the right-(left-) moving electrons  close to the Fermi points as
	\begin{align}
	\Psi(x) = \sum\limits_{\eta =\pm} \Psi_{\eta}(x) e^{i\eta k_F x},
	\label{eqn:Psi}
	\end{align}	
	where $k_F$ is the Fermi momentum\label{def:k_F}.	Here and below we set $\hbar =1$.
	 The effective Hamiltonian describing the electrons in the topological nanowire near the Fermi points reads~\cite{Gangadharaiah}	
	\begin{align}
	\begin{split}
	H_{\mathrm{e}} &= \int dx \sum\limits_{\eta =\pm} \Psi^\dag_\eta(x)\left( -iv_F \eta \partial_x\right) \Psi_\eta(x)\\
	&\hspace{50pt}- \int dx \ \Delta \left(i\Psi_+^\dag(x)\Psi_-^\dag(x) + \rm {H.c.} \right),
	\end{split}
	\label{eqn:linearized-Hamiltonian}
	\end{align}
	where $v_F$ is the Fermi velocity, $\Delta>0$ denotes the superconducting gap $\Delta(k)$ at $k\approx k_F$ in the TSC, 
	and the integrals over $x$, here and also below, are over the domain $x\in[ 0, L]$. In the following we assume that the length $L$ is much greater than the coherence length $\xi=v_F/\Delta$, $L\gg \xi$. The Bogolyubov-de Gennes (BdG) equations corresponding to the Hamiltonian in Eq.~(\ref{eqn:linearized-Hamiltonian}) can be written in the following form:
	\begin{align}
	\left(-iv_F\eta_z\partial_x + \Delta \eta_y \tau_x\right)  \Phi_\alpha = \varepsilon_\alpha \Phi_\alpha,
	\label{eqn:BdG}
	\end{align}
	where the index $\alpha$ labels the eigenstates, $\Phi = \begin{pmatrix}\phi_{+},&\phi_{-},&\bar{\phi}_{+},&\bar{\phi}_{-}\end{pmatrix}^T$  is a spinor whose components describe right- and left-moving electrons ($\phi_{+}$ and $\phi_{-}$, respectively) and their hole counterparts ($\bar{\phi}_{\eta}$), $\varepsilon_\alpha$ is the corresponding eigenenergy. The Pauli matrices $\eta_{x,y,z}$ and $\tau_{x,y,z}$\label{def:eta_xyz}\label{def:tau_xyz} act on the space of states with different chiralities and on the particle-hole space, respectively. The field operators in Heisenberg representation then satisfy
	\begin{align}
	&\Psi(x,t) =\; \sum\limits_{\alpha, \eta}\left[c_\alpha \phi_{\eta,\alpha}(x)e^{i\eta k_Fx  -i \varepsilon_\alpha t}
	\right. \label{eqn:Heisenberg1} \\
	&\hspace{80pt}\left.+\;c_\alpha^\dag \bar{\phi}_{\eta,\alpha}(x) e^{-i\eta k_Fx  +i \varepsilon_\alpha t}\right],\nonumber 
\\
	&\Psi_{\eta}(x,t) =\; \sum\limits_\alpha \left[  c_\alpha \phi_{\eta, \alpha}(x) e^{-i\varepsilon_\alpha t} + c_\alpha^\dag \bar{\phi}_{-\eta, \alpha}(x) e^{i\varepsilon_\alpha t}\right],\label{eqn:Heisenberg2}
	\end{align}
	where $c_\alpha$ are fermionic quasiparticle annihilation operators. The normalization condition for the spinor $\Phi_\alpha$ reads
	\begin{align}
	\frac{1}{2}|\Phi|^2 = \frac{1}{2}\sum\limits_{\eta=\pm}\int dx \left( |\phi_{\eta}|^2 + |\bar{\phi}_{\eta}|^2\right) = 1.
	\end{align}
	We use vanishing boundary conditions for the field $\Psi$ at the end points $x=0$ and $x=L$:
	\begin{align}
	\Psi(x=0) = \Psi(x=L) = 0,
	\end{align}
	which can be rewritten in terms of $\phi_{\eta}$ as
	\begin{align}
		&\sum\limits_{\eta=\pm} \phi_{\eta}(x=0) =\sum\limits_{\eta=\pm} \phi_{\eta}(x=L) e^{i\eta k_F L} = 0 \, ,
			\label{eqn:linearized-BCs1}\\	
	&\sum\limits_{\eta=\pm} \bar{\phi}_{\eta}(x=0) =\sum\limits_{\eta=\pm} \bar{\phi}_{\eta}(x=L) e^{-i\eta k_F L} = 0\, .
	\label{eqn:linearized-BCs2}
	\end{align}
	We note that the model given by Hamiltonian $H_{\mathrm{e}}$ [see Eq.~(\ref{eqn:linearized-Hamiltonian})], with boundary conditions defined by Eqs.~(\ref{eqn:linearized-BCs1})--(\ref{eqn:linearized-BCs2}), can be used both as an effective low-energy model for the Kitaev chain~\cite{Kitaev2001}  or for electrons in Rashba nanowire~\cite{Alicea2010, Lutchyn2010, Oreg2010, Mourik2012,  Das2012, Deng2012, Sticlet2012, Churchill2013, Rokhinson2012,  Klinovaja2012, Chevallier2012, San-Jose2012, Dominguez2012, Terhal2012, Klinovaja2012a, Prada2012, DeGottardi2013, Thakurathi2013, Maier2014,  Escribano2017, Prada2017, Ptok2017, Kobialka2018, DeMoor2018} in the regime of spin-polarized electrons when Zeeman field is much greater than the proximity-induced superconducting gap~\cite{Gangadharaiah} (we consider a more general lattice model for spinful topological Rashba nanowires in Sec.~\ref{sec:Rashba}). The hallmark of the 1D TSC is the presence of MBSs. The BdG equations defined in Eq.~(\ref{eqn:BdG}), with the boundary conditions given by ~Eqs.~(\ref{eqn:linearized-BCs1})--(\ref{eqn:linearized-BCs2}), have two subgap solutions $\Phi_{\pm}(x) = \left[\Phi_L(x) \pm i\Phi_R(x)\right]/\sqrt{2}$ with exponentially small eigenenergies,
	\begin{align}
	\varepsilon_{\pm} = \pm \Delta e^{-L/\xi}\sin \left(k_F L\right),
	\label{eqn:MBS-energy} 
	\end{align}
	and where $\Phi_{L(R)}(x)$\label{def:Phi_LR} is the spinor corresponding to the left (right) Majorana mode (see Appendix~\ref{app:BdG} for details). The wavefunctions corresponding to the MBSs have the following form:
	\begin{align}
	\langle 0|\Psi(x)|\pm\rangle &= \sum\limits_{\eta=\pm} \phi_{\eta,\pm}(x)e^{i\eta k_F x} = \sqrt{\frac{2}{\xi}}\sin (k_F x)e^{-x/\xi} \nonumber\\&\pm \sqrt{\frac{2}{\xi}}\sin [k_F (L-x)]e^{-(L-x)/\xi}.
	\end{align}	
For a finite length $L$, there are discrete bulk modes $\Phi_n(x)$\label{def:Phi_n} (labeled by the index $n=\pm 1, \pm 2, \dots$) with eigenenergies $\varepsilon_n$\label{def:varepsilon_n}. We note that particle-hole symmetry implies that $\varepsilon_n = -\varepsilon_{-n}$. 
	
	The Hamiltonian of the phonon bath reads
	\begin{align}
	H_{\mathrm{ph}} = \sum\limits_q \Omega_q b_q^\dag b_q,
	\end{align}
	where $b_q$ is the annihilation operator for a phonon with momentum $q$, and $\Omega_q$ is the phonon energy. For 1D acoustic phonons, we assume $\Omega_q = c_s q$ where $c_s$ is a speed of sound.
	The electron-phonon interaction for 1D phonons can be described by the following Hamiltonian\cite{AGD, Mahan2000}:
	\begin{align}
	H_{\mathrm{e-ph}} = \gOne\int dx\; \sum\limits_{\eta =\pm}\Psi_\eta^\dag(x)\Psi_\eta(x)\varphi(x),
	\label{eqn:H_e-ph}
	\end{align}
	where $\varphi(x) = \sum\limits_q \left(c_sq/2L\right)^{1/2}\left(b_q e^{iqx} + b_q^\dag e^{-iqx}\right)$ is the phonon field\cite{AGD}, $\gOne$ is the electron-phonon coupling strength, which can be estimated for acoustic phonons as $\gOne \simeq C_{ac}/(c_s\sqrt{\rho_A a^2})$~\cite{Mahan2000, Cleland2002}, where $C_{ac}$\label{def:Cac} is the acoustic deformation potential coupling constant, $\rho_A$\label{def:rho_A} is the atomic mass density, $c_s$\label{def:c_s} is the speed of sound, and $a$\label{def:a} is the lattice constant.

	\begin{figure}
	\includegraphics[width=\columnwidth]{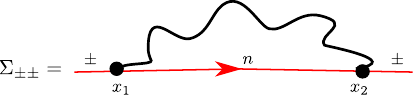}		
	\caption{ Feynman diagram for the self-energy $\Sigma_{\pm\pm}$ corresponding to the process where  a phonon (curly black line) promotes the system from  a bound state $\Phi_{\pm}$ to a bulk state $n$ (straight red  line).}
	\label{fig:feynman}
\end{figure}	
We treat the electron-phonon interaction as a perturbation using the Keldysh formalism~\cite{Keldysh1965diagram, Kamenev09} in  first order in $\gOne^2$. The absorbed virtual or thermal phonon can promote the electron subsystem from the bound state $\Phi_{\pm}$ to one of the bulk states $\Phi_n$ (see Fig.~\ref{fig:feynman}). This process  shifts the energies of the bound states, so that the new energies are $\varepsilon_{\pm} + \delta \varepsilon_{\pm}$. Particle-hole symmetry implies $\delta \varepsilon_- = -\delta \varepsilon_+$, and the 
correction to the energy splitting between the MBSs is $|2\delta \varepsilon_+|$.  Since the electron stays in the bound state only for a finite time $\tau_0$ \label{def:tau_0} before being promoted to a bulk state, the energy levels of the MBSs are broadened, with the level broadening $\gamma$ being inverse proportional to $\tau_0$ according to the uncertainty principle, $\gamma = \hbar/\tau_0$.   The shift in energy $\delta \varepsilon_{\pm}$ \label{def:varepsilon_pm} and the broadening $\gamma$\label{def:Gamma} of the MBS can be related to the real and the imaginary parts of the retarded self-energy $\Sigma^R_{\pm\pm}(\varepsilon)$:
\begin{align} 
&\delta \varepsilon_{\pm} =\; \mathrm{Re}\; \Sigma_{\pm\pm}^R(\varepsilon_\pm), 
\label{eqn:delta-epsilon} 
&\gamma =\; \mathrm{Im}\; \Sigma_{\pm\pm}^R(\varepsilon_\pm), 
\end{align} 
where the retarded self-energy (see Fig.~\ref{fig:feynman}) reads in leading order~\cite{Aseev2018},
\begin{align} 
\Sigma^R_{\pm\pm}(\varepsilon) = \sum\limits_n\int dx_1 dx_2\ \rho_{\pm, n}(x_1) W^R_n(x_1, x_2, \varepsilon) \rho_{n,\pm}(x_2). 
\end{align} 
Here, 
the sum goes over both positive and negative $n$, $\rho_{m,n}(x) =\Phi^\dag_m(x) \tau_z\Phi_n(x)$, and the effective interaction $W^R_n(x_1,x_2, \varepsilon)$ is given by a convolution of electron and phonon Green functions~\cite{Aseev2018}:
\begin{align}
\begin{split}
&W^R_n(x_1, x_2, \varepsilon) = \gOne^2\int \frac{d\omega }{2\pi}\left[G^R_n(\varepsilon - \omega)\right.\\
&\left. \times D^K(x_1-x_2, \omega) + G^K_n(\varepsilon - \omega) D^A(x_1-x_2, \omega)\right],
\end{split}
\label{eqn:effective-interaction}
\end{align}
where $G^{R(A)}_n(\varepsilon) = (\varepsilon - \varepsilon_n \pm i0^+)^{-1}$\label{def:GRA} are retarded (advanced) Green functions for electrons in the eigenbasis, $G^K_n(\varepsilon) = -2\pi i \delta(\varepsilon-\varepsilon_n)\tanh[\varepsilon_n/(2T)]$\label{def:GK} is their Keldysh counterpart, and $D^{R(A),K}$ denote the corresponding phonon Green functions. We  note that $W^R_n(x_1, x_2, \varepsilon)=W^R_n(x_1-x_2, \varepsilon)$, reflecting the fact that we assumed translation invariance for the phonon modes, i.e., we assume that the boundary effects of the finite-sized nanowire on the phonons are negligible (in contrast to the electron system). In the following we assume that electron and phonon subsystems are in thermal equilibrium. However, a generalization to the case when temperatures of electron and phonon subsystems are different can be done straightforwardly.
Finally, we note that the inverse of the broadening, $1/\gamma$, can be interpreted as the lifetime of the states $\Phi_{\pm}(x)$ formed by the MBSs, i.e. the characterstic time it takes until these quasiparticle states change their occupation due to the interaction with  phonons.
 
 We further note that particle-hole symmetry implies that $\rho_{m,n}(x) = \rho_{-n,-m}(x)$, $\rho_{L,n}(x) = \rho_{-n,L}(x)$, and $\rho_{R,n}(x) = - \rho_{-n,R}(x)$.
It is convenient to rewrite Eq.~(\ref{eqn:delta-epsilon}) in the basis of $\Phi_{L, R}$ instead of $\Phi_{\pm}$:
\begin{multline} 
\delta \varepsilon_+ = \int dx dx'\;\\ \times i\sum_n\rho_{L,n}(x) \mathrm{Re} \left[W^R_n(x-x', \varepsilon\approx 0)\right] \rho_{n,R}(x').
\end{multline} 
Similarly, the broadening $\gamma = \Sigma_{++}^R(\varepsilon\approx 0) = \Sigma_{--}^R(\varepsilon\approx 0)$ can be expressed as a sum of two contributions from the opposite edges, $\gamma = \gamma_L + \gamma_R$, where
\begin{align}
\begin{split}
\gamma_{L(R)} &=\; \int dxdx'\sum\limits_{n} \rho_{L(R),n}(x)\\
&\times\mathrm{Im}\left[W^R_n(x-x', \varepsilon\approx 0)\right]\rho_{n,L(R)}(x').
\end{split}
\end{align}
Here and below we disregard an exponentially small difference between the non-perturbed energies of the MBSs, $\varepsilon_{\pm}$, and for the calculation of both the self-energy $\Sigma^R$ and the effective interaction $W^R_n$, the approximation $\varepsilon\approx 0$ is used. 
	
	\section{Interaction with 1D acoustic phonons \label{sec:1D-acoustic}}	
In this section, we consider effects of interaction between electrons in topological nanowire and 1D acoustic phonons with a linear spectrum $\Omega_q = c_s q$, where $c_s$ is a speed of sound. The retarded (advanced) phonon Green function reads:
	\begin{align}
	D^{R(A)}(x,\omega) = \int \frac{dq}{2\pi} \frac{\Omega_q^2}{\omega^2 - \Omega_q^2 \pm i0^+} e^{iqx}.
	\end{align} 
	The Keldysh counterpart can be obtained using fluctuation-dissipation theorem, $D^K(\omega) = \left[D^R(\omega)-D^A(\omega)\right]\coth\left(\omega/2T\right)$.
	
	Using Eq.~(\ref{eqn:effective-interaction}) and performing integration over phonon frequency $\omega$, we obtain that the effective interaction $W^R$ can be represented as a sum of two contributions (see Appendix~\ref{app:eff-interaction} for details): $W^R_{v}$ describing the processes including only virtual transitions and $W^R_a$ those including absorption of phonons, $W^R = W^R_v + W^R_a$: 
	\begin{align}
	&W^R_{v,n}(x) = \gOne^2\frac{ \pi \varepsilon_nT^2}{c_s}\, \sum\limits_{k=1}^{\infty }\, \frac{k e^{- \frac{2\pi k T |x|}{c_s}}}{\varepsilon_n^2 + 4k^2 \pi^2 T^2} ,
	\label{eqn:W-virtual}\\
	&W^R_{a,n}(x) = i\gOne^2\frac{\varepsilon_n}{4c_s}\left[\tanh \left(\frac{\varepsilon_n}{2T}\right)-\coth \left( \frac{\varepsilon_n}{2T}\right)   \right] e^{-i\frac{|x|\varepsilon_n}{c_s}}
	\label{eqn:W-absorption}. 
	\end{align}
	
	The  term $W^R_{v,n}$ contributes to the real part of the self-energy determining the splitting of the MBSs, while  $W^R_{a,n}$ contributes both to the imaginary and the real part of the self-energy, i.e., both to the splitting and to the broadening of the MBSs. In the case when the electron and phonon subsystems are equilibrated  independently and are at different temperatures, $T_e$ and $T_{ph}$, respectively, the contribution from the virtual processes is governed by the temperature of the phonon bath $T_{ph}$, while the absorption term takes the following form:
		\begin{align}
		&W^R_{a,n}(x) = i\gOne^2\frac{\varepsilon_n}{4c_s}\left[\tanh \left(\frac{\varepsilon_n}{2T_e}\right) -\coth \left( \frac{\varepsilon_n}{2T_{ph}}\right)  \right] \nonumber \\
		& \hspace{150pt} \times e^{-i\frac{|x|\varepsilon_n}{c_s}}.
		\end{align}
We note that if the temperature of the phonon bath is finite, $T_{ph} >0$, the term $W^R_{a,n}$ does not vanish even for zero electron temperature, $T_e=0$.

	\subsection{Zero temperature $T=0$}
	\begin{figure}
		\includegraphics[width=\columnwidth]{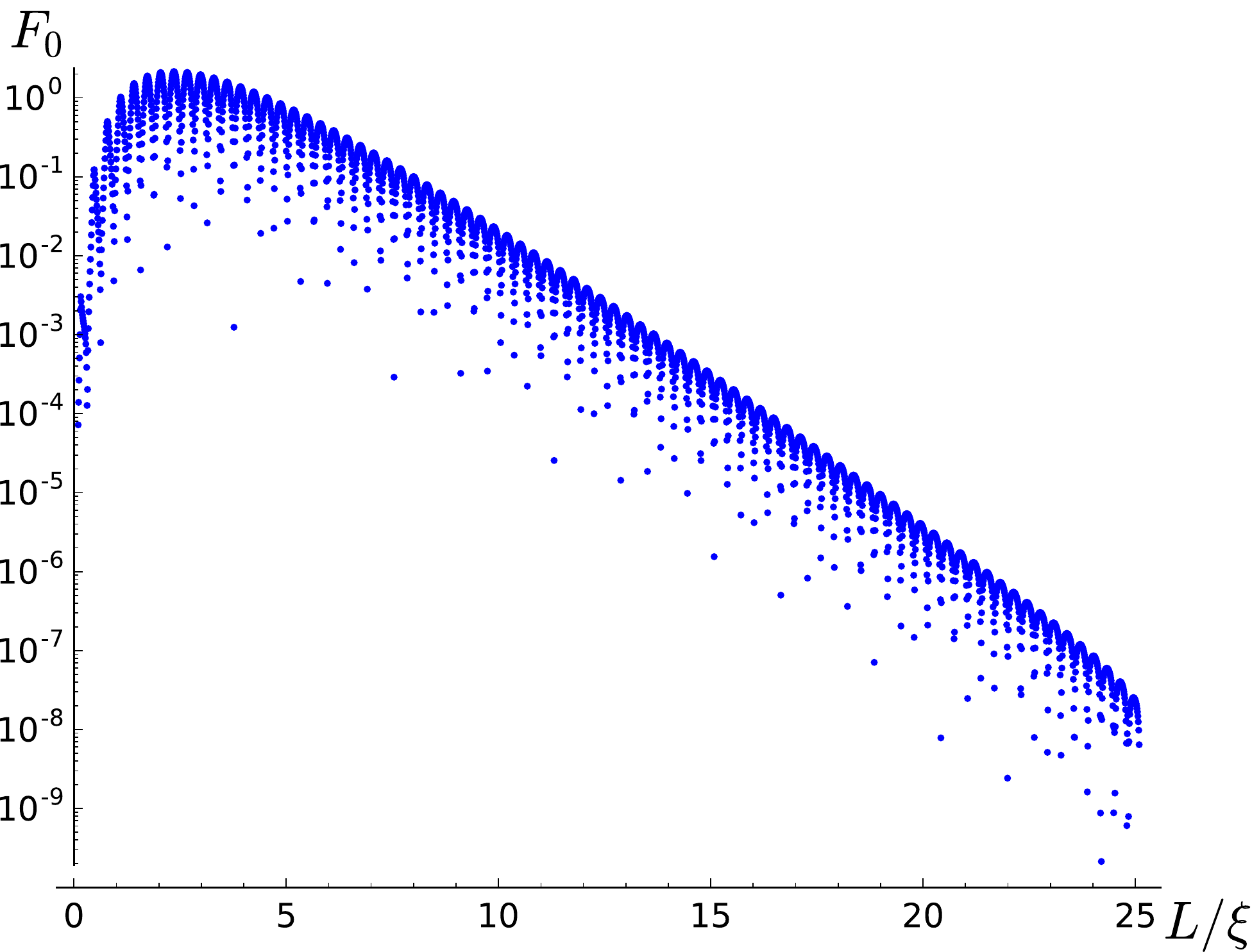}
		\caption{Plot of the dimensionless function $|F_0(L)|$ defined in Eq. (\ref{eqn:F0}) that describes the  $L$-dependence  of the energy splitting $\delta \varepsilon_+$ of the MBSs [see Eq. (\ref{eqn:result-acoustic})]. $|F_0(L)|$ shows oscillatory behaviour with a period determined by $k_F$ and an amplitude exponentially decaying as $\exp(-L/\xi)$. Here, we took $k_F = 10/\xi$. }
		\label{fig:zeroT-decay}
	\end{figure}
	
	In the limit $T \to 0$, the contribution $W^R_{a,n}\propto \exp\left\{-\Delta/T\right\}$ due to absorption of real phonons vanishes, and the contribution from the exchange of virtual phonons (see Appendix~\ref{app:eff-interaction} for details) reduces to
	\begin{align}
	W^R_{v,n}(T=0,x) = \frac{\gOne^2 c_s }{4\pi\varepsilon_n x^2}.
	\end{align}
	Thus, the imaginary part of the effective interaction vanishes at zero temperature, so that the broadening $\gamma$ tends to zero. The correction to the energy of the MBS, 
	$\delta \varepsilon_+$, takes the following form:
	\begin{align}
	 \delta \varepsilon_+&=\; \frac{\gOne^2 c_s}{4\pi \Delta L^2} F_0(L),
	\label{eqn:result-acoustic}
	\end{align}
	where $F_0(L)$ is given by
	\begin{align}
	\begin{split}
	F_0(L) &=\; \sum\limits_n \int dx dx' \; \rho_{L,n}(x) \frac{\Delta L^2}{\varepsilon_n (x-x')^2} \rho_{n,R}(x')\\
	&\approx \sum\limits_n \frac{\Delta}{\varepsilon_n}  P_{L,n} P_{n,R}.
	\end{split}
	\label{eqn:F0}
	\end{align}
	Here, $P_{L,n} = \int dx\; \rho_{L,n}(x)$ and $P_{n,R} = \int dx\; \left[i\rho_{n,R}(x)\right]$. In the last equality we used the fact that $\rho_{L,n}(x)$ and $\rho_{n,R}(x)$ decay exponentially when $x\gtrsim \xi$ and $L-x \gtrsim \xi$, respectively, so that the main contribution to the integral comes from $x\approx0$, $x' \approx L$.  The summation over bulk states labeled by $n$  results in an exponential decay, $\delta \varepsilon_+ \propto \exp\left(-L/\xi\right)$ if $L\gg \xi$, as shown in Fig.~\ref{fig:zeroT-decay}. This can be easily understood since the process shown in Fig.~\ref{fig:feynman} involves  virtual transitions between the bound states and the bulk states, and the corresponding lifetime of a virtual excitation in the bulk band is finite, being of order $\hbar/\Delta$, and, thus, cannot propagate further than $\hbar v_F/\Delta=\xi $, the coherence length.
	
	\subsection{Finite temperatures $T>0$}
	
		\begin{figure}
		\includegraphics[width=\columnwidth]{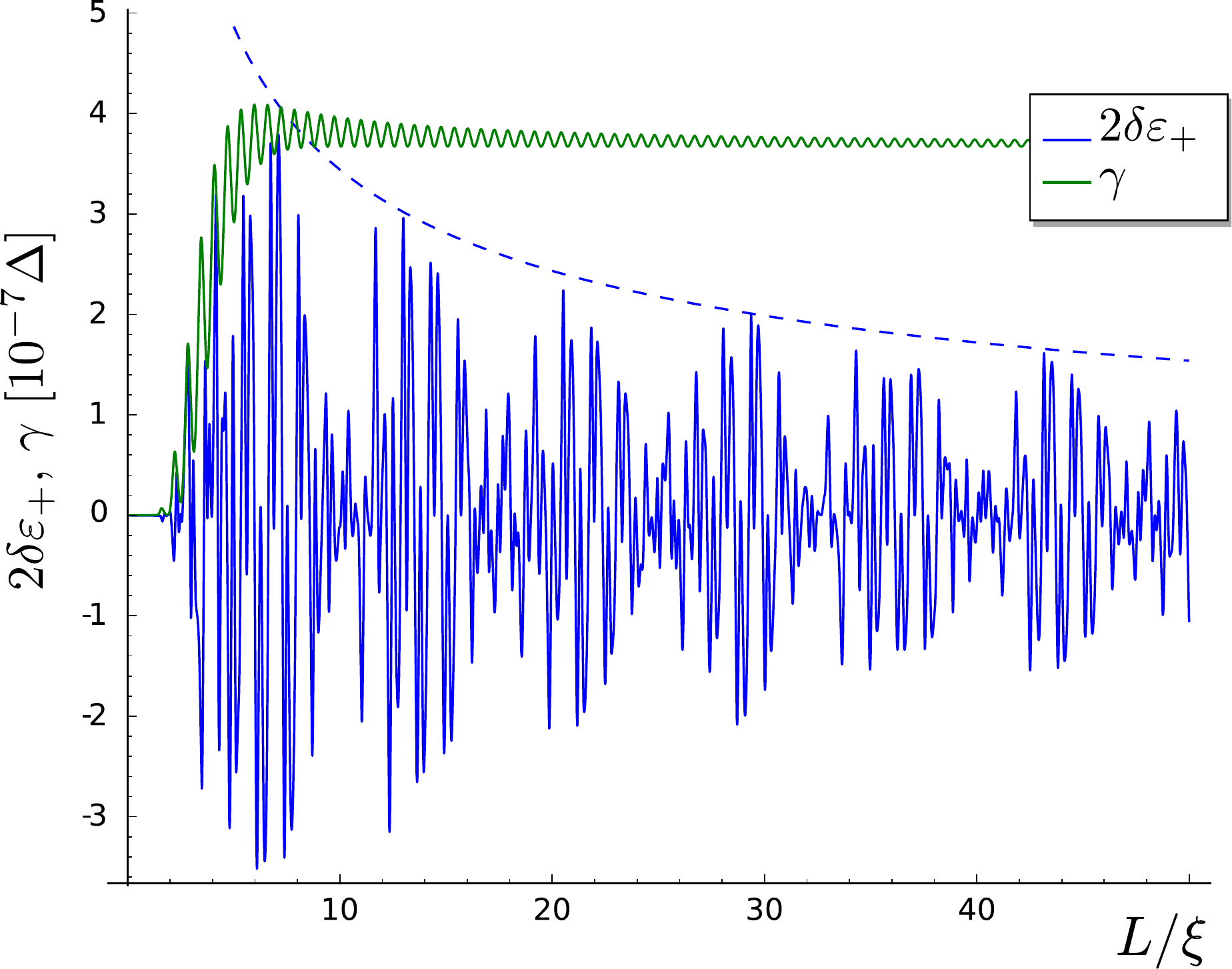}
		\caption{Energy splitting  $2\delta \varepsilon_+$ (blue soild line) and broadening $\gamma$ (green line) of MBSs caused by interaction with 1D acoustic phonons [see Eqs.~(\ref{eqn:acoustic-dE-finiteT}) and (\ref{eqn:acoustic-Gamma-finiteT})]. 
		Dashed blue line shows analytical estimation for $2\delta \varepsilon_+$ given by Eq.~(\ref{eqn:acoustic-estimation}). The broadening $\gamma$ saturates at a finite value as the length $L$ of the nanowire grows, while the energy splitting $2\delta \varepsilon_+$ oscillates with a characteristic oscillation scale governed by both Fermi wavelength $2\pi /k_F$ and phonon wavelength $c_s/\Delta$. 
		The amplitude  of $2\delta \varepsilon_+$ decays as a power-law $\propto L^{-1/2}$  with increasing length. We took $(\Delta,v_F,c_s, C_{ac},a) = (0.2\;\mathrm{meV}, 4.6\cdot10^{4}\;\mathrm{m/s}, 4.4\cdot 10^{3}\;\mathrm{m/s}, 5\;\mathrm{eV}, 5\;\mathrm{\AA}$), 
		$T=0.1\Delta$, $k_F = 5/\xi$ for numerical estimations.}
		\label{fig:eta_acoustic}
	\end{figure}

At finite temperatures, the leading contribution to the sum in Eq.~(\ref{eqn:W-virtual}) comes from $k=1$, and the effective interaction decays exponentially on the scale of the phonon thermal length $l_{ph} = \hbar c_s/(2\pi T)$, i.e. $W^R_{v,n} \propto \exp\left(-|x|/l_{ph}\right)$. Thus, in the following we disregard the contribution from the virtual phonons. The contribution from the processes involving absorptions of phonons [see ~Eq.~(\ref{eqn:W-absorption})] gives rise 
	to the splitting of the MBSs given by 
\begin{multline} 
\delta\varepsilon_+ = \frac{\gOne^2}{4\hbar c_s} \sum_n \frac{\varepsilon_n}{\sinh\left(\varepsilon_n/T\right)}\\ \times \int dxdx'\; i\rho_{L,n}(x) \rho_{n,R}(x')\sin\left(\frac{\varepsilon_n |x-x'|}{c_s}\right)\\\approx \frac{\gOne^2}{4\hbar c_s} \sum_n \frac{\varepsilon_n}{\sinh\left(\varepsilon_n/T\right)}\int dx \rho_{L,n}(x) e^{-i\varepsilon_n x/c_s}\\
\times \int dx'\;\left[i\rho_{n,R}(x')\right] e^{i\varepsilon_n x'/c_s},
\label{eqn:acoustic-dE-finiteT}
\end{multline} 
and to the broadening $\gamma = \gamma_{L} + \gamma_{R}$ given by
	\begin{multline}
		\gamma_{L(R)}(T) = \frac{\gOne^2}{4\hbar c_s} \sum_n \frac{\varepsilon_n}{\sinh\left(\varepsilon_n/T\right)}\times\\ \int dxdx'\; \rho_{L(R),n}(x) \rho_{n,L(R)}(x')\cos\left(\frac{\varepsilon_n (x-x')}{c_s}\right)\\
		= \frac{\gOne^2}{4\hbar c_s} \sum_n \frac{\varepsilon_n}{\sinh\left(\varepsilon_n/T\right)}\left|\int \rho_{L(R),n}(x) e^{i\varepsilon_n x/c_s}\right|^2
		\label{eqn:acoustic-Gamma-finiteT}.
	\end{multline}
	
	The integration over $x$ is performed in Appendix~\ref{app:matrix}. The resulting dependences of splitting and broadening on the length $L$ are shown in Fig.~\ref{fig:eta_acoustic}. If $L\gg \xi$, we find that the broadening saturates at a finite value:
	\begin{align}
	\gamma(L\gg \xi) \sim \frac{\gOne^2}{\xi} \frac{c_s}{v_F} e^{-\Delta/T}.
	\end{align}
	The energy shift  $\delta \varepsilon_+$ shows oscillations governed both by the Fermi wavelength $2\pi/k_F$ and phonon wavelength of order $c_s/\Delta$. 
	The amplitude decays as a power-law in $L$: 
	\begin{align}
	\delta \varepsilon_+ \propto \frac{\gOne^2}{\sqrt{L\xi}}\frac{c_s}{v_F}e^{-\Delta/T}.
	\label{eqn:acoustic-estimation}
	\end{align}
	This power-law behavior
	is a result of the coherent emission and absorption of thermal phonons. 
	At zero temperature, both the energy shift $\delta \varepsilon_+$ and the broadening $\gamma$ vanish in agreement with the results of the previous section. For finite temperature we note that although the energy shift of the MBSs and, thus, the splitting between the two MBSs scales as a power-law rather than exponentially with increasing $L$, an experimental observation of the splitting can be challenging since  $2\delta \varepsilon_+$ remains always smaller than the broadening $\gamma$, which does not scale with $L$. 
	
	The result that $\gamma$ does not depend on $L$ (for large $L$) is not surprising and in line with earlier results~\cite{Schmidt2012}: the lifetime of a fermion level formed by the left and right MBS can be affected by local couplings alone such that the right and left MBS can contribute independently to the linewidth. In stark contrast, the energy splitting $\delta \varepsilon_+$ of the MBSs (and thus the lifting of the degeneracy) is  possible only via non-local couplings involving both the right and the left MBS simultaneously. Indeed, the two MBSs get  hybridized because the bulk quasiparticle that gets excited by absorbing a thermal phonon around, say, the left MBS, can propagate
	over the entire distance $L$ to the right MBS before it decays again by emitting a phonon. The remarkable consequence of this phonon-induced correlation is that  the splitting $\delta \varepsilon_+$ decays only as a power law and not exponentionally in $L$.
	In this sense, thermal phonons lift the topological protection of  the groundstate degeneracy of the TSC at any finite temperature. This is still the case
	for vanishing electron-temperature of the TSC, $T_e=0$, as long as the temperature of the phonon system is finite, $T_{ph}>0$.
	The exponential (topological) protection of the TSC is only established when the entire system, electrons and phonons, is  at strictly zero temperature, i.e. $T=T_e=T_{ph}=0$.

\section{Interaction with confined phonon branches with Van Hove singularities \label{sec:quasi-1D}}

Now we assume that the higher dimensional phonons are confined in transverse direction so that their motion in direction perpendicular to the nanowire is quantized, and the motion along the nanowire can  still be described by a continuum momentum $q$ along the $x$-axis. The phonon energy spectrum reads
\begin{align}
\Omega_{q, j}^2 = c_s^2 q^2 + \Omega_{j}^2,
\end{align} 
where $j$ \label{def:j} labels the transverse phonon branch, $\Omega_{j}$ \label{def:Omega_0j} is the energy of the bottom of the $j$-th branch.  For simplicity we consider the two lowest branches, so that the spectrum of phonons of the lowest branch is linear, $\Omega_{0} = 0$, $\Omega_{q, 0} = c_s q$, and the energy of the bottom of the second branch $\Omega_{1}$ is close to the gap $\Delta$, which is possible if the phonons are confined in the transverse direction on the scale of $c_s/\Delta$. The phonons are now described by the phonon fields $\varphi_j(x)$, where $j=0,1$, and the Hamiltonian describing the electron-phonon interaction can be written as
\begin{align}
H_{\mathrm{e-ph}} = \sum\limits_{j=0,1;\, \eta = \pm} g_j \int dx\; \Psi^\dag_\eta(x) \Psi_\eta(x) \varphi_j(x),
\end{align}  
where $g_j$ is the interaction strength between the electrons and the $j$-th phonon branch.
In  leading order of the perturbation theory, both phonon branches  contribute to the self-energy independently.
The contributions of the lowest branch with $j=0$ to the energy splitting and broadening of the MBSs are similar to those considered in the previous section. 

\subsection{Effect of Van Hove singularity} 
\label{VHSsingularity}

Next we focus on the contribution from the phonon branch with $j=1$. First, we note that the density of states for these phonons contains a Van Hove singularity (VHS) at zero momentum since $\left(d\Omega_{q,1}/dq\right)_{q=0} = 0$ (see App.~\ref{app:VHS} for discussion of VHSs in nanowires).  In the following we consider how this singularity affects the MBSs.

Using Eq.~(\ref{eqn:effective-interaction}) and integrating over the phonon frequency $\omega$ and momentum $q$, we obtain the following contribution to the effective interaction from the absorption of thermal phonons (see Appendix~\ref{app:eff-interaction} for details):
\begin{multline}
W^R_{a,n}(x) =
\begin{cases}
 \frac{g_1^2}{8c_s} \frac{\varepsilon_n^2
   e^{- |x|\sqrt{\Omega_{1}^2 - \varepsilon_n^2}/c_s }}{\sinh \left(\varepsilon_n/T\right) \sqrt{\Omega_1^2 - \varepsilon_n^2}},&\varepsilon_n < \Omega_1,\\
   \\
  -i\frac{g_1^2}{8c_s} \frac{\varepsilon_n^2 e^{-i|x|\sqrt{ \varepsilon_n^2 - \Omega_1^2}/c_s }}{\sinh \left(\varepsilon_n/T\right) \sqrt{\varepsilon_n^2 - \Omega_1^2}},&\varepsilon_n > \Omega_1.
 \end{cases}
 \label{eqn:WR_VHS}
\end{multline}
Here, we 
have disregarded contributions from  virtual transitions since they are short-ranged in $L$, similarly to the previous section. The singularity at the resonance condition $\varepsilon_n = \Omega_1$ is due to the VHS at the band bottom of the phonon branch. 
Finally, we obtain the shift of the MBS energy $\delta \varepsilon_+= \delta \varepsilon_{+}^< + \delta \varepsilon_{+}^>$ and the broadening $\gamma = \gamma_R + \gamma_L$,  
\begin{widetext}
	\begin{align}
	\delta \varepsilon_{+}^< &=\; \frac{g_1^2}{4\hbar c_s}\int dxdx'\;\mathrm{Re}\;\left\{ \sum\limits_{n,\varepsilon_n< \Omega_1} i\rho_{L,n}(x)\rho_{n,R}(x')
	\frac{\varepsilon_n^2 \exp\left[-|x'-x|\sqrt{\Omega_{1}^2 - \varepsilon_n^2}/c_s \right]}{\sinh \left(\varepsilon_n/T\right) \sqrt{(\Omega_1+i\Gamma)^2 - \varepsilon_n^2}} \right\},\label{eqn:VHS-dE}\\
		\delta \varepsilon_{+}^> &=\; \frac{g_1^2}{4\hbar c_s}\int dxdx'\;\mathrm{Re}\;\left\{\sum\limits_{n,\varepsilon_n> \Omega_1} i\rho_{L,n}(x)\rho_{n,R}(x')  \frac{\varepsilon_n^2 \sin\left[|x'-x|\sqrt{\varepsilon_n^2 - \Omega_1^2}/c_s\right]}{\sinh \left(\varepsilon_n/T\right) \sqrt{\varepsilon_n^2 - (\Omega_1+i\Gamma)^2}}  \right\}, \label{eqn:VHS-dE2} \\
	\gamma_{L(R)} &=\; \frac{g_1^2}{4\hbar c_s}\int dxdx'\;\mathrm{Re}\;\left\{ \sum\limits_{n, \varepsilon_n>\Omega_1} \rho_{L(R),n}(x)\rho_{n,L(R)}(x') \frac{\varepsilon_n^2\cos\left[(x-x')\sqrt{\varepsilon_n^2 - \Omega_1^2}/c_s\right]}{\sinh\left(\varepsilon_n/T\right) \sqrt{\varepsilon_n^2 - (\Omega_1+i\Gamma)^2}}\right\}.
	\label{eqn:VHS-Gamma}
	\end{align}
\end{widetext}

\begin{figure}[t]
	\includegraphics[width=0.95\linewidth]{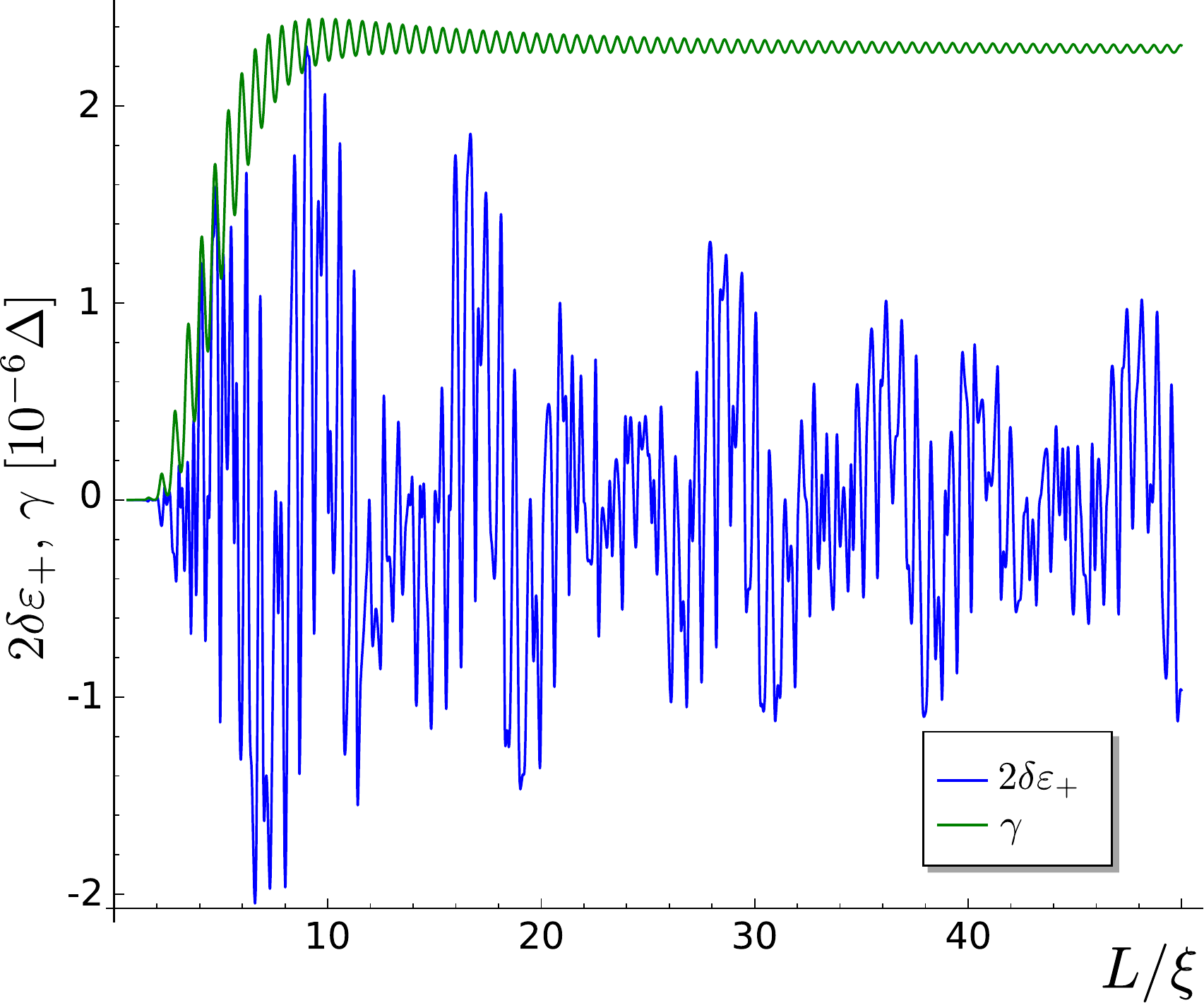}
	\caption{Energy splitting  $2\delta \varepsilon_+$ (blue line) and broadening $\gamma$ (green line) [see Eqs.~(\ref{eqn:VHS-dE})--(\ref{eqn:VHS-Gamma})] of the MBSs due to the second branch of confined phonons in case when the band bottom of this branch,  $\Omega_1$, is less than the gap $\Delta$, $\Omega_1 = 0.9\Delta$. In this case,
		neither the broadening $\gamma$, nor the energy splitting $2\delta \varepsilon_+$ exhibit resonance peaks. Comparing with Fig.~\ref{fig:eta_acoustic} one can see that the dependence of $\gamma$ and $2\delta \varepsilon_+$ on $L$ is qualitatively the same as for purely 1D acoustic phonons without VHS. We took $g_1 = \gOne$, and the phenomenological relaxation rate $\Gamma=0.01\;\mathrm{\mu eV}$. The other parameters are  the same as in Fig.~\ref{fig:eta_acoustic}. }
	\label{fig:VHS_vs_L_ES09}
\end{figure}

\begin{figure}[t]
	\includegraphics[width=0.95\linewidth]{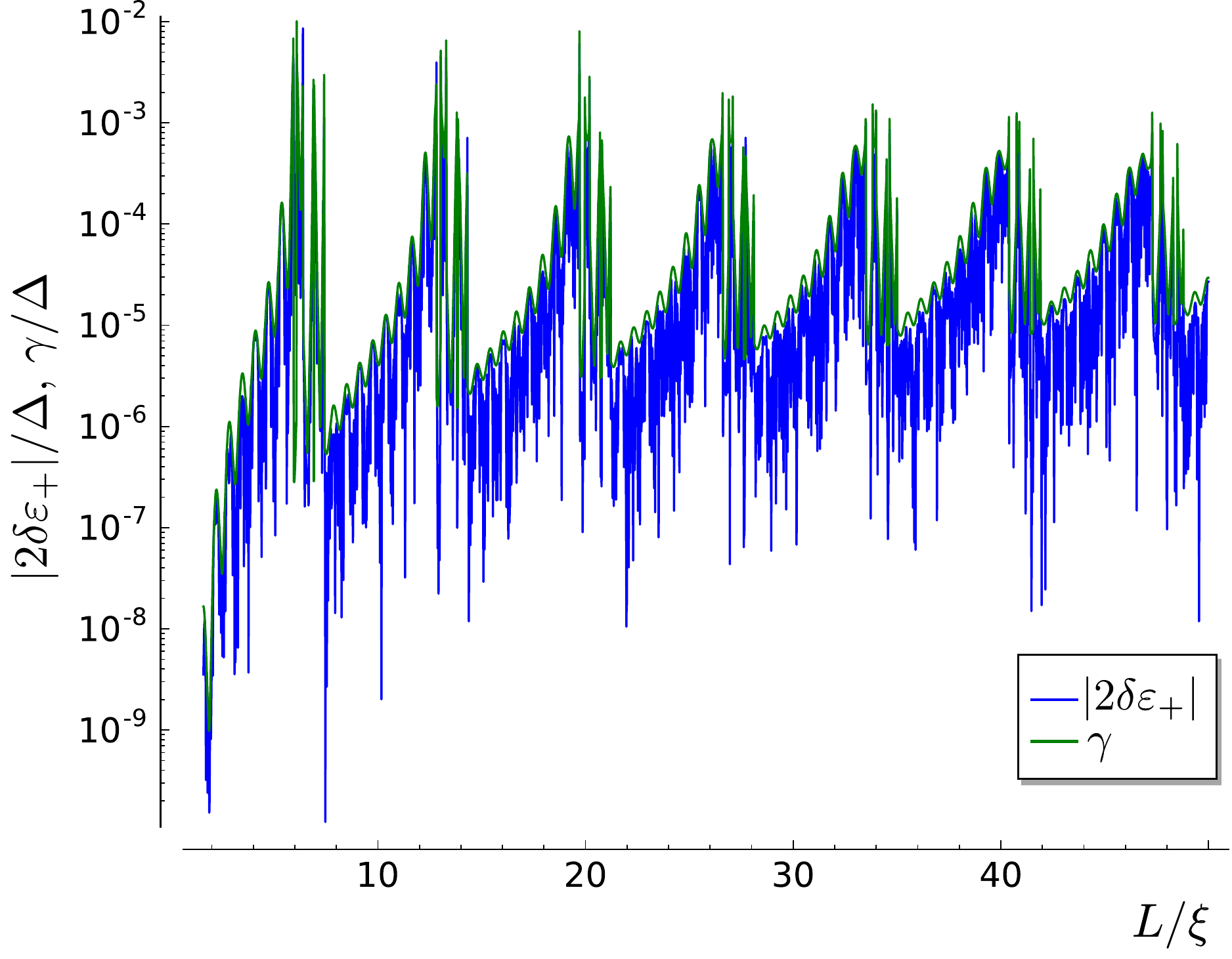}
	\caption{The same as Fig.~\ref{fig:VHS_vs_L_ES09} but for the case when $\Omega_1$ slightly exceeds the gap $\Delta$, $\Omega_1 = 1.1\Delta$. If the nanowire length $L$ is 
		varied, each time
		$\Omega_1$ coincides with  energies of the bulk quasiparticle states, both the broadening $\gamma$ and  splitting $2\delta \varepsilon_+$ are enhanced, manifesting resonance peaks. The splitting  $2\delta \varepsilon_+$ remains significant and of order of $\gamma$ even for $L\gg \xi$.  The other parameters are  the same as in Fig.~\ref{fig:VHS_vs_L_ES09}. }
	\label{fig:VHS_vs_L}
\end{figure}

In order to regularize the VHS singularities at $\varepsilon_n = \Omega_1$ we introduced a phenomenological parameter $\Gamma$, which describes finite broadening for the phonon energy induced by, e.g., anharmonic effects (giving rise to phonon-phonon interaction). We note that the broadening $\Gamma$ due to anharmonic effects can be estimated as $\Gamma(T) \sim \Gamma_A\left\{1 + 2/\left[\exp\left(-\Omega_1/T\right) - 1\right]\right\}$,
where we use $\Gamma_A \sim 0.01\;\mathrm{\mu eV}$ as an estimate \cite{Klemens1966,Balkanski1983,Menendez1984}. In principle, it is also reasonable to take into account the  broadening $\gamma$ of the electron levels caused by the electron-phonon interaction and include it into $\Gamma$ self-consistently. However, we checked that our results  depend only weakly on the exact value of $\Gamma$ (see also below).

We note that the sums over $n$ in Eqs.~(\ref{eqn:VHS-dE})--(\ref{eqn:VHS-Gamma}) cannot be replaced by integrals since the summands contain rapidly varying exponents and sine factors. Therefore, in order to get finite results, we perform the summation numerically. However, it is possible to get analytical estimates when $\Omega_1$ is close to one of the eigenvalues $\varepsilon_n$ and the term with the corresponding $n$ dominates. The contribution $\delta \varepsilon_{+}^<$ originates from transition to the quasiparticle bulk states with energies $\varepsilon_n$ less than the energy of the bottom of the phonon branch $\Omega_1$ due to interaction via decaying phonon modes, while $\delta \varepsilon_{+}^>$ describes the splitting caused by transitions to quasiparticle states with energies $\varepsilon_n$ larger than $\Omega_1$ due to interaction with propagating phonons with momentum $\sqrt{\Omega_1^2 - \varepsilon_n^2}/c_s$. For long nanowires, the term with a specific eigenvalue $\varepsilon_n$ close to $\Omega_1$ gives the dominant contribution to $\delta \varepsilon_{+}^<$. In case of such a resonance, $ \Omega_1 - \varepsilon_n < c_s^2/(2L^2 \Omega_1)$, one can estimate 
\begin{align}
\delta \varepsilon_{+}^< \sim \frac{g_1^2}{c_s} \frac{\Omega_1^{3/2}e^{-\Omega_1/T}}{\sqrt{ \max\{\Gamma,\Omega_1 - \varepsilon_n \}}}\frac{\xi}{L}.
\end{align}
If the bottom of the phonon branch $\Omega_1>\Delta$ does not coincide with any quasiparticle energy $\varepsilon_n$, then $\Omega_1-\varepsilon_n\sim v_F/L$, and, hence, the contribution  $\delta \varepsilon_{+}^<$ vanishes subexponentially with increasing length $L$:
\begin{align}
 \delta\varepsilon_{+}^< \sim \frac{g_1^2}{c_s} \Omega_1^{3/2}  \frac{\sqrt{v_F}\xi}{L^{3/2}} e^{-\sqrt{2\Omega_1L v_F}/c_s -\Omega_1/T}.
\end{align}

Similarly, if $\Omega_1$ is close to $\varepsilon_n$, so that  $\Omega_1 < \varepsilon_n$, $\varepsilon_n - \Omega_1<c_s^2/(2L^2 \Omega_1)$, the contribution $\gamma_{res}$  to $\gamma_{L(R)}$ from the resonant term can be estimated as
\begin{align}
\gamma_{res} \approx \frac{g_1^2}{c_s} \frac{\Omega_1^{3/2}e^{-\Omega_1/T}}{\sqrt{ \max\{\Gamma, \varepsilon_n - \Omega_1 \}}}\frac{\xi}{L}.
\end{align}
We note that, although the resonant contributions to splitting and the broadening are of the same order, the resonance condition for the splitting requires the bulk mode with energy $\varepsilon_n$ to be less than $\Omega_1$, $\varepsilon_n\lesssim
\Omega_1$, whereas the resonance condition for the broadening requires $\varepsilon_n\gtrsim\Omega_1$.

The contribution $\gamma_{0} = \gamma_{L(R)} - \gamma_{res}$ from the rest of the quasiparticle states with $\varepsilon_n>\Omega_1$ can be estimated in the thermodynamic limit $L/\xi \to \infty$
(see Appendix~\ref{app:estimations}) as
\begin{align}
&\gamma_0 = g_1^2 \sum\limits_{n;\, \varepsilon_n>\Omega_1} \frac{\varepsilon_n^2 e^{-\varepsilon_n/T}}{2\Omega_1\max\{\varepsilon_n - \Omega_1, \Gamma\}} \nonumber
\\& \hspace{110pt}\times
 \int dx\; \rho_{Ln}(x)\rho_{nL}(x).
\end{align}
Also, in this limit, summation over $n$ can be replaced by integration, and one obtains the estimate (see Appendix~\ref{app:estimations})
\begin{align}
\gamma_0 \sim \frac{g_1^2}{\pi c_s}\frac{\Delta^2}{\sqrt{\Omega_1^2 - \Delta^2}}e^{-\Omega_1/T}\ln \left( \frac{T}{\Gamma}\right). 
\end{align}
We  note that the contribution to the splitting $\delta \varepsilon_{+}^>$ oscillates with the period being governed by $k_F$ and $c_s/\Delta$. The amplitude of oscillations remains smaller than the broadening, i.e. $|\delta \varepsilon_{+}^>| \lesssim  \gamma$ (see Appendix \ref{app:estimations}).

\begin{figure}
	\includegraphics[width=0.9\columnwidth]{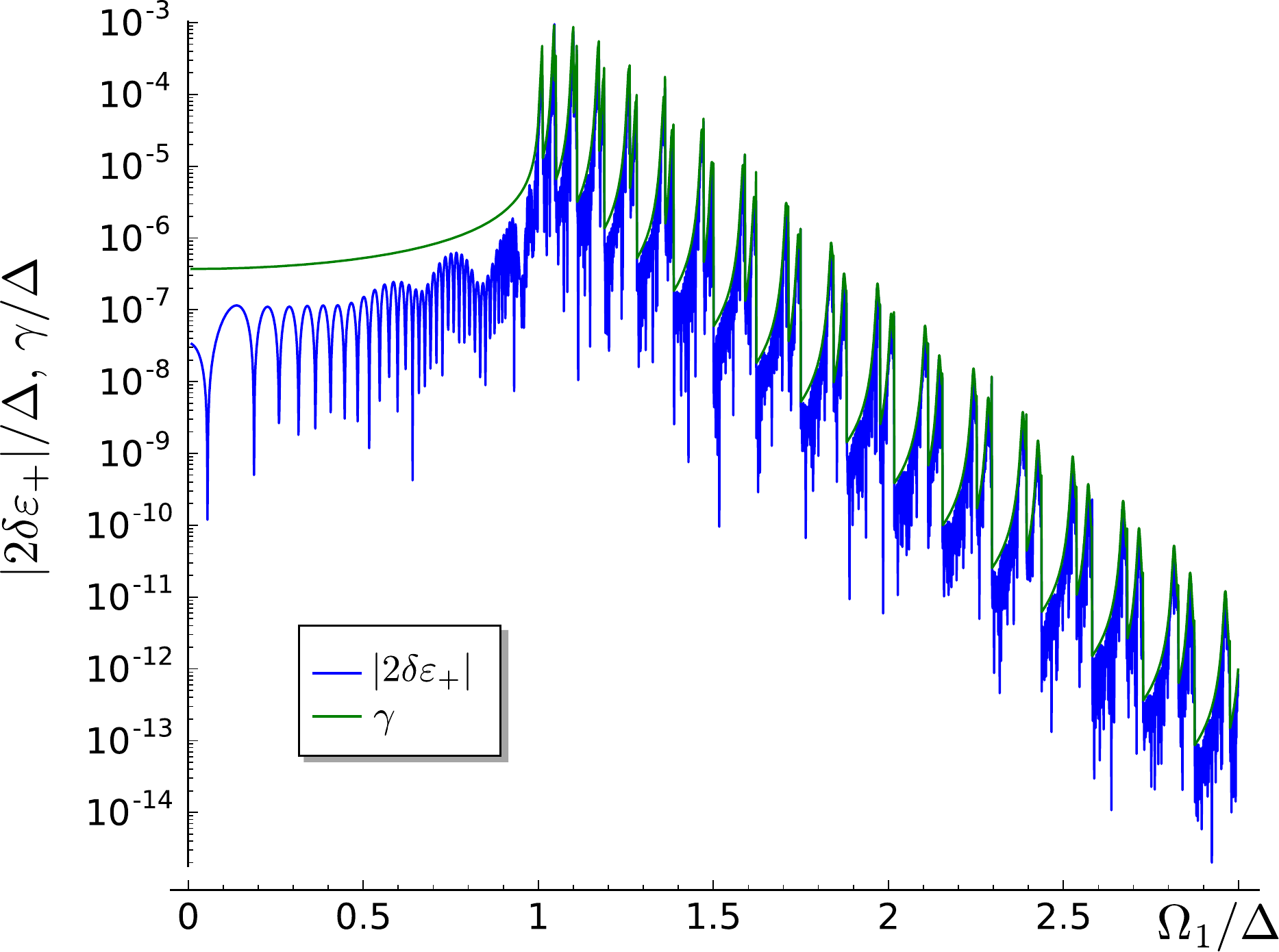}
	\caption{The same as Fig.~\ref{fig:VHS_vs_L} but as function of the energy $\Omega_1$ of the band bottom of the phonon branch. If $\Omega_1 < \Delta$ the broadening $\gamma$ and the amplitude of the energy splitting $\delta \varepsilon_+$ depend weakly on $\Omega_1$. The absorption of phonons and, thus, $\delta\varepsilon_+$ and $\gamma$, are enhanced as $\Omega_1$ gets close to $\Delta$ due to the VHS in the phonon branch, and the resonance peaks correspond to the case when $\Omega_1$ coincides with an energy of a bulk quasiparticle mode $\varepsilon_n$. At increasing $\Omega_1$, both  $\gamma$ and $\delta \varepsilon_+$ decay exponentially in temperature, $\delta \varepsilon_+, \gamma \propto \exp\left\{-\Omega_1/T\right\}$. We took $L=20\xi$. The other parameters are  the same as in Fig.~\ref{fig:VHS_vs_L}}
	\label{fig:VHS_vs_Omega}
\end{figure}

\begin{figure*}[]
	\includegraphics[width=0.9\linewidth]{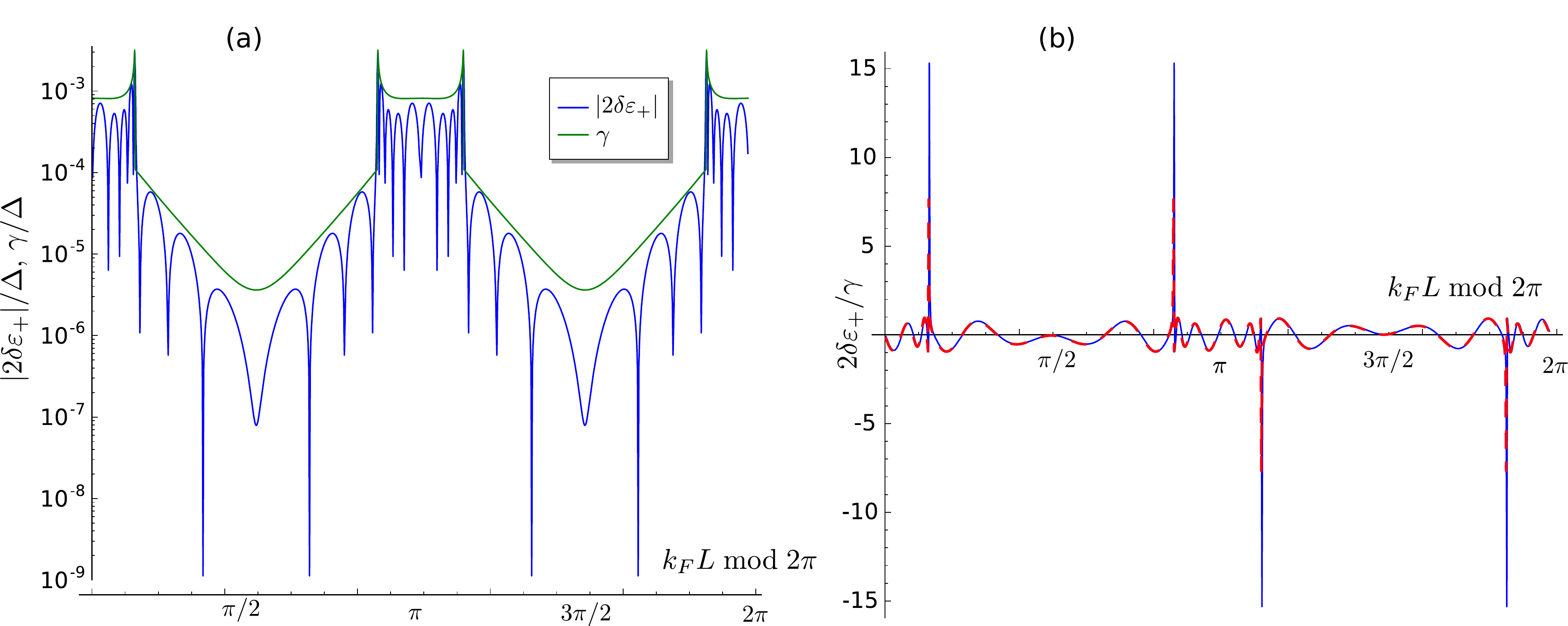}
	\caption{(a) The energy splitting $2\delta \varepsilon_+$ (blue line) and broadening $\gamma$ (green line) of the MBSs  and (b) the ratio $\delta \varepsilon_+/\gamma$ as function of $k_F$ for $\Gamma = 0.01\;\mathrm{\mu eV}$ (blue solid line) and $\Gamma = 1\;\mathrm{\mu eV}$ (red dashed line). By tuning $k_F$ (for example, by gates) one can achieve a resonance when the band bottom of the phonon branch $\Omega_1$ coincides with an energy of one of the quasiparticle bulk modes. The resonant absorption of phonons boosts both $\delta\varepsilon_+$ and $\gamma$. Around the resonance, the energy splitting $\delta\varepsilon_+$ can largely exceed the broadening $\gamma$. Thus, a fine-tuning of $k_F$ makes it possible to observe the phonon-induced energy splitting of the MBSs. The height of the resonance peaks is determined by the value of $\Gamma$, however, this dependence is logarithmically weak. We took $L=20\xi$. The other parameters are the same as in Fig.~\ref{fig:VHS_vs_L}.}
	\label{fig:VHS_vs_kF}
\end{figure*}

We plot the resulting energy splitting $|2\delta \varepsilon_+|$ [see Eq.~(\ref{eqn:VHS-dE})--(\ref{eqn:VHS-dE2})] and  broadening $\gamma$ [see Eq.~(\ref{eqn:VHS-Gamma})] as functions of the length $L$ (see Fig.~\ref{fig:VHS_vs_L_ES09} and Fig.~\ref{fig:VHS_vs_L}), $\Omega_1$ (see Fig.~\ref{fig:VHS_vs_Omega}), and $k_F$ (see Fig.~\ref{fig:VHS_vs_kF}). If $\Omega_1$ is below the gap $\Delta$, then the broadening and splitting, 
which are caused by interaction with the second phonon branch  $j=1$, remain of the same order of magnitude as for the phonon lowest branch $j=0$ (see Fig.~\ref{fig:VHS_vs_L_ES09}). However, the absorption of phonons becomes strongly enhanced as the VHS at energy $\Omega_1$ coincides with energies of bulk quasiparticle states, resulting in pronounced resonance peaks in $\delta \varepsilon_+$ and $\gamma$ 
(see Fig.~\ref{fig:VHS_vs_L}). If $\Omega_1$ becomes larger than the gap $\Delta$, the absorption of phonons decays exponentially in temperature, yielding $ \delta \varepsilon_+, \gamma \propto \exp\left\{-\Omega_1/T \right\}$ and manifesting resonance peaks when $\Omega_1$ coincides with an energy of a bulk quasiparticle mode $\varepsilon_n$ (see Fig.~\ref{fig:VHS_vs_Omega}). 

It is important to note that the interaction of the quasiparticles with the phonon lowest branch $j=0$ results in a splitting $2\delta \varepsilon_+$ that is significantly less than the broadening $\gamma$ (see Fig.~\ref{fig:eta_acoustic}). However, the interaction with a second phonon branch that has a VHS at zero momentum results in splittings $2\delta \varepsilon_+$  close to $\gamma$. Moreover, close to the resonances, when the contribution $\delta \varepsilon_{+}^<$ dominates, the energy shift $\delta \varepsilon_+$ can be by an order of magnitude larger than the broadening (see Fig.~\ref{fig:VHS_vs_kF}), so that the energy splitting between the MBSs becomes observable. However, in this case a fine-tuning may be required. It is also important to point out that the dependence of the results on the exact value of the phenomenological regularization parameter $\Gamma$ is logarithmically weak [see Fig.~\ref{fig:VHS_vs_kF} (b)]. We also note that if the bottom of the second phonon branch is close to the gap, $\Omega_1\approx \Delta$, the absorption processes for this mode dominate over those for the phonon lowest branch  $j=0$. For example, if both interaction strengths are taken the same, $g_1 = g_2$, the broadening and  splitting for the higher phonon band $j=1$ can be by a factor of $10^2$ (or even $10^4$ close to the resonance) larger than those due to interaction with the phonon lowest branch $j=0$. 

We finally remark that in realistic settings it is difficult to avoid such resonances between one of the bulk states and the VHS singularities. For example, 
if one of the MBS is shifted by changing the length of the topological section (i.e. the length $L$) with the help of voltage gates, one unavoidably hits such a resonance, see Fig. \ref{fig:VHS_vs_L}. Generally, even keeping $L$ constant, the bulk levels could themselves fluctuate due to various external, even local, perturbations resulting in resonances with $\Omega_1$. We also note that our model could be easily generalized  to the case of several phonon branches, each of which is characterized by its own 
VHS energy $\Omega_n$. Each of these branches results in   enhanced MBS energy splitting and  broadening when one of the bulk levels of the TSC happens to get into resonance with one of the VHS $\Omega_n$.

\subsection{Spinful lattice model for topological Rashba nanowires \label{sec:Rashba}}

\begin{figure*}[t]
	\includegraphics[width=0.9\linewidth]{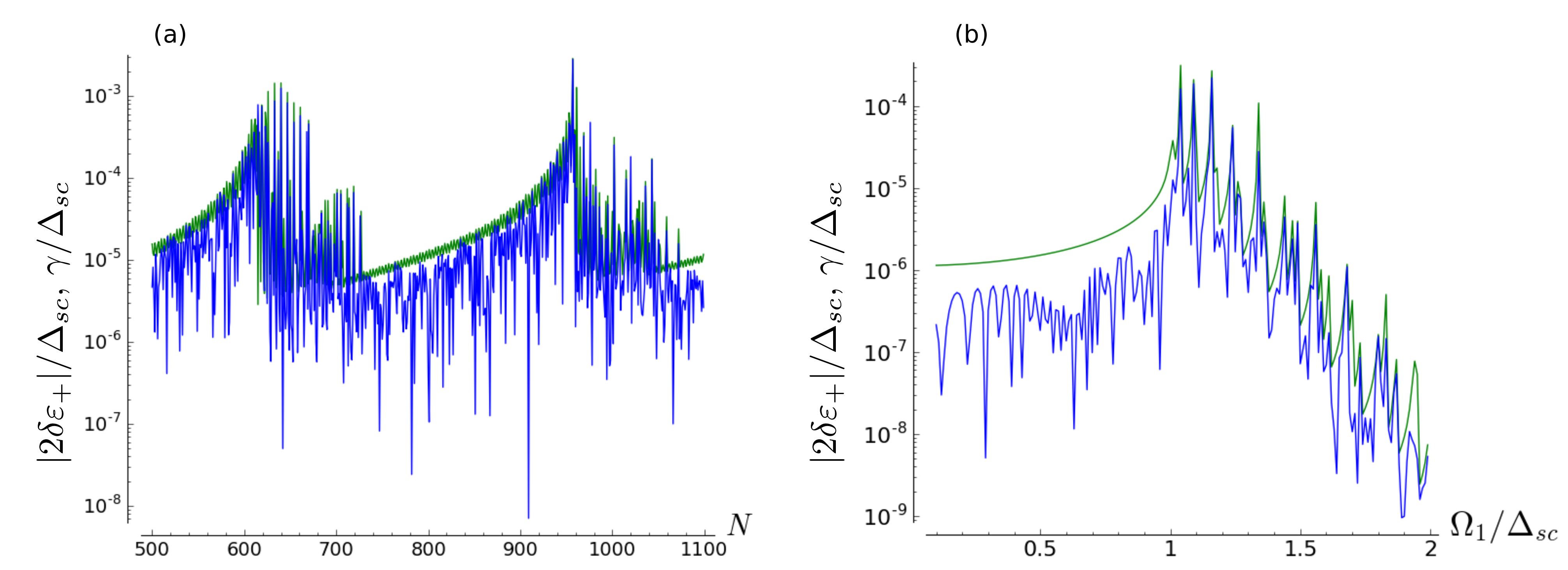}
	\caption{Energy splitting $2\delta \varepsilon_+$ (blue line) and broadening $\gamma$ (green line) of the MBSs as functions of  (a) number of sites $N$  and (b) bottom of phonon branch $\Omega_1$ for the spinful lattice model [see Eq.~(\ref{eqn:Hamiltonian_Rashba})] of a topological Rashba nanowire. A higher mode of confined phonons interacts with electrons causing the splitting and broadening of the MBSs. The behavior is qualitatively similar to the case of the low-energy continuum model (see Fig.~\ref{fig:VHS_vs_L} and Fig.~\ref{fig:VHS_vs_Omega}).  If the length $L=Na_0$ of the 1D TSC is fine-tuned so that  $\Omega_1$ coincides with one of the energy levels  of the bulk quasiparticle states, both the broadening $\gamma$ and the energy splitting $2\delta \varepsilon_+$ show pronounced resonance peaks. Importantly, the energy shift $\varepsilon_+$ remains significant and of order of $\gamma$ even in long samples, $L\gg \xi$. The parameters chosen are $(E_{so}, \Delta_{sc}, \Delta_Z, \Omega_1,\mu) = (0.6, 0.2, 0.4, 0.22,0)\; \mathrm{meV}$, corresponding to $t=10\mathrm{meV}$ and coherence length $\xi = 50a_0$. The rest of the parameters are the same as for Fig.~\ref{fig:VHS_vs_L}.}
	\label{fig:Rashba1_ab}
\end{figure*}

The low-energy continuum model for spinless electrons defined by Eq.~(\ref{eqn:linearized-Hamiltonian}) is justified if the Fermi energy $\mu$ is much larger than the quasiparticle gap $\Delta$ and temperature $T$.
In order to complement our considerations, we consider a more realistic model describing a topological Rashba nanowire. Namely, we consider a spinful single-band nanowire with a proximity-induced superconducting gap $\Delta_{sc}$. The Rashba spin-orbit interaction (SOI) of  strength $\alpha_R$ sets the spin quantization axis to be perpendicular to the nanowire and corresponds to the SOI energy $E_{so} = m_0\alpha_R^2/(2\hbar^2)$, where $m_0$ is the effective electron mass. A magnetic field (corresponding to the Zeeman energy $\Delta_Z$) is applied along the nanowire.
The tight-binding Hamiltonian describing the nanowire reads~\cite{Rainis2013}
\begin{align}
&H =\; \sum\limits_{j,s', s}
c^\dag_{s',j+1}\left[-t\delta_{s' s} -\frac{i}{2a_0}\alpha_R\sigma^y_{s's} \right] c_{s,j} + H.c. \nonumber\\
&+\sum\limits_{j,s,'s} c^\dag_{s',j}\left[2t\delta_{s's} - \mu \delta_{s's} + \Delta_Z\sigma^x_{s's}  \right] c_{s,j} \nonumber
\\
&+\;\sum\limits_{j} 	\Delta_{sc} \left(c^\dag_{\uparrow, j}
c^\dag_{\downarrow,j} + H.c.\right), 
\label{eqn:Hamiltonian_Rashba}
\end{align}	
where $t=\hbar^2/(2m_0a_0^2)$ is the hopping amplitude, $a_0$ is the lattice constant of the tight-binding model, $c_{s,j}$ annihilates an electron with spin $s$ at site $j$ with coordinate $x_j$, $\mu$  is a uniform chemical potential (we assume that $\mu=0$ corresponds to the energy $E_{so}$ measured from the bottom of the band), and $\sigma^{x,y}$ are the Pauli matrices. In the following we assume that the Zeeman energy is larger than the superconducting proximity gap, $\Delta_Z > \sqrt{\Delta_{sc}^2+\mu^2}$, which brings the system into the topological phase with one MBS localized at each end of the chain~\cite{Rainis2013}.
The energy splitting and level broadening of the MBSs due to interaction with a phonon mode with 
VHS can be calculated by using expressions similar to Eqs.~(\ref{eqn:VHS-dE})--(\ref{eqn:VHS-Gamma}) with integrations over the coordinate $x$ replaced by summations over sites $x_j$: 

\begin{figure*}[t]
	\includegraphics[width=0.95\linewidth]{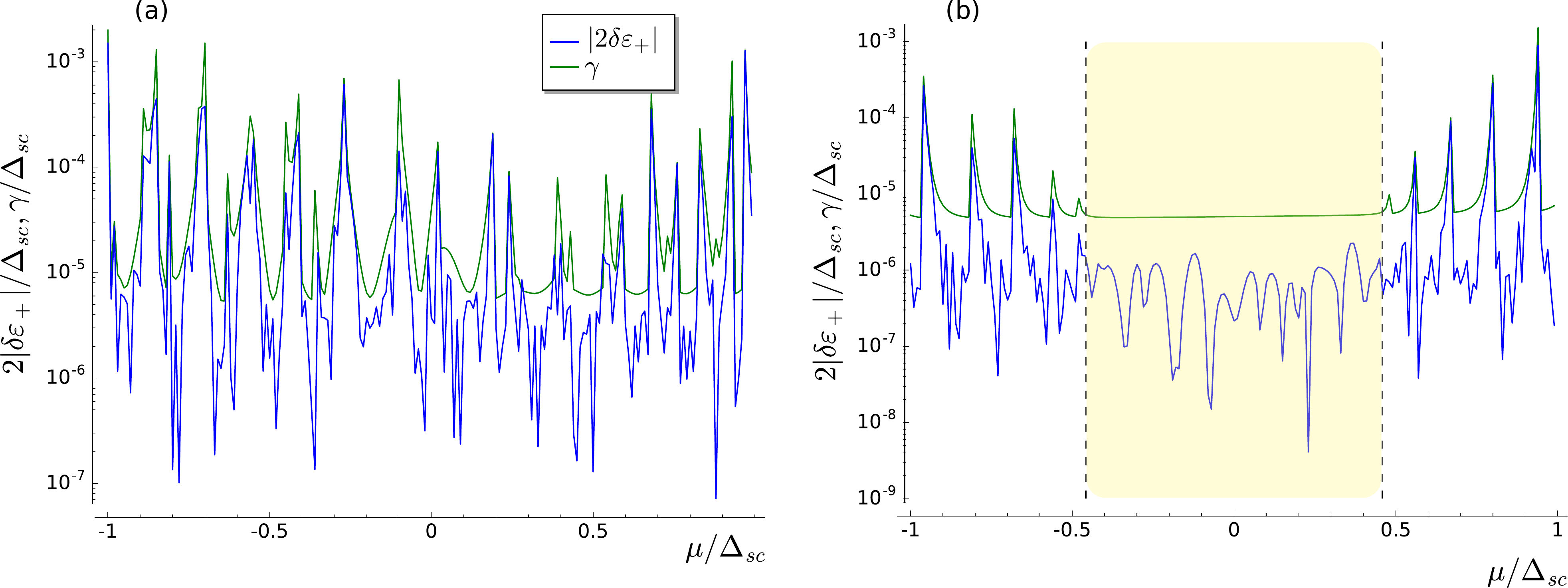}
	\caption{Energy splitting $2\delta \varepsilon_+$ (blue line) and broadening $\gamma$ (green line) of the MBSs
		as functions  of chemical potential $\mu$ for VHS energies (a) $\Omega_1 = 1.1\Delta_{sc}$  and (b)  $\Omega_1 = 0.9\Delta_{sc}$  calculated numerically for the spinful lattice model of a topological Rashba nanowire, see Eq. (\ref{eqn:Hamiltonian_Rashba}). By tuning $\mu$, one can achieve the resonance as the bottom of the phonon band coincides with an energy of one of the quasiparticle  bulk eigenstates. The resonant absorption of phonons results in the growth of both $\delta \varepsilon_+$ and $\gamma$ by several orders of magnitude. 
		(b) If $\Omega_1 < |\Delta_Z - \Delta_{sc}|$, the quasiparticle gap is larger than $\Omega_1$ for small values of chemical potential $|\mu|$ (shaded yellow region). Thus, the broadening $\gamma$ does not depend on $\mu$, while the splitting $|2\delta \varepsilon_+|$ remains smaller than the broadening $\gamma$. As $|\mu|$ increases, the topological gap $\Delta = |\Delta_Z-\sqrt{\Delta_{sc}^2 - \mu^2}|$ decreases. At some point, $\Omega_1$ becomes greater than $\Delta$, and both broadening $\gamma$ and  energy splitting $|2\delta \varepsilon_+|$ show resonant behavior, such that, depending on $\mu$, the energy splitting can become comparable to the broadening, and, thus, can be observed experimentally. We took $L=20\xi$, the rest of parameters are the same as for Fig.~\ref{fig:Rashba1_ab}.}
	\label{fig:VHS_vs_mu_Rashba}
\end{figure*}

\begin{widetext}
	\begin{align}
	\delta \varepsilon_{+}^< &=\; \frac{g_1^2}{4c_s}a_0^2\sum\limits_{j, j'}\;\mathrm{Re}\;\left\{ \sum\limits_{n,\varepsilon_n< \Omega_1} i\rho_{L,n}(x_j)\rho_{n,R}(x_{j'})
	\frac{\varepsilon_n^2 \exp\left[-|x_{j'}-x_j|\sqrt{\Omega_{1}^2 - \varepsilon_n^2}/c_s \right]}{\sinh \left(\varepsilon_n/T\right) \sqrt{(\Omega_1+i\Gamma)^2 - \varepsilon_n^2}} \right\},\\
	\delta \varepsilon_{+}^> &=\; \frac{g_1^2}{4c_s}a_0^2\sum\limits_{j,j'}\;\mathrm{Re}\;\left\{\sum\limits_{n,\varepsilon_n> \Omega_1} i\rho_{L,n}(x_j)\rho_{n,R}(x_{j'})  \frac{\varepsilon_n^2 \sin\left[|x_{j'}-x_j|\sqrt{\varepsilon_n^2 - \Omega_1^2}/c_s\right]}{\sinh \left(\varepsilon_n/T\right) \sqrt{\varepsilon_n^2 - (\Omega_1+i\Gamma)^2}} 
	\right\}, \\
	\gamma_{L(R)} &=\; \frac{g_1^2}{4c_s}a_0^2\sum\limits_{j,j'}\mathrm{Re}\;\left\{ \sum\limits_{n, \varepsilon_n>\Omega_1} \rho_{L(R),n}(x_j)\rho_{n,L(R)}(x_{j'}) \frac{\varepsilon_n^2\cos\left[(x_j-x_{j'})\sqrt{\varepsilon_n^2 - \Omega_1^2}/c_s\right]}{\sinh\left(\varepsilon_n/T\right) \sqrt{\varepsilon_n^2 - (\Omega_1+i\Gamma)^2}}\right\}
	\label{eqn:VHS-Gamma-Rashba},
	\end{align}
\end{widetext}
where $\rho_{L(R),n}(x) = \Phi_{L(R)}^\dag(x) \tau_z \Phi_n(x)$ and $\Phi_n$ ($\Phi_{L,R}$) is the eigenspinor of  $H$ [see  Eq.~(\ref{eqn:Hamiltonian_Rashba})]  written in the basis $(c_{\uparrow j},  c_{\downarrow j}, c_{\uparrow j}^\dag,  c_{\downarrow j}^\dag)$.

The resulting splitting $|2\delta \varepsilon_+|$ and level broadening $\gamma$ are plotted as functions of number of sites $N$, bottom of the phonon branch $\Omega_1$, and chemical potential $\mu$ for the spinful lattice model [see Eq.~(\ref{eqn:Hamiltonian_Rashba})] in Figs.~\ref{fig:Rashba1_ab} and \ref{fig:VHS_vs_mu_Rashba}. Comparing with the corresponding plots obtained for the low-energy continuum model (see Figs.~\ref{fig:VHS_vs_L_ES09}--\ref{fig:VHS_vs_Omega}), we conclude that the results obtained in these two models are qualitatively very similar. Thus, the same remarks made above (see end of Subsec. \ref{VHSsingularity}) regarding hitting accidentally such resonances when changing $N$ during manipulations of the MBSs apply here as well.

\subsection{Charge and spin of the  MBSs induced by phonons} 

\begin{figure*}[]
	\includegraphics[width=\linewidth]{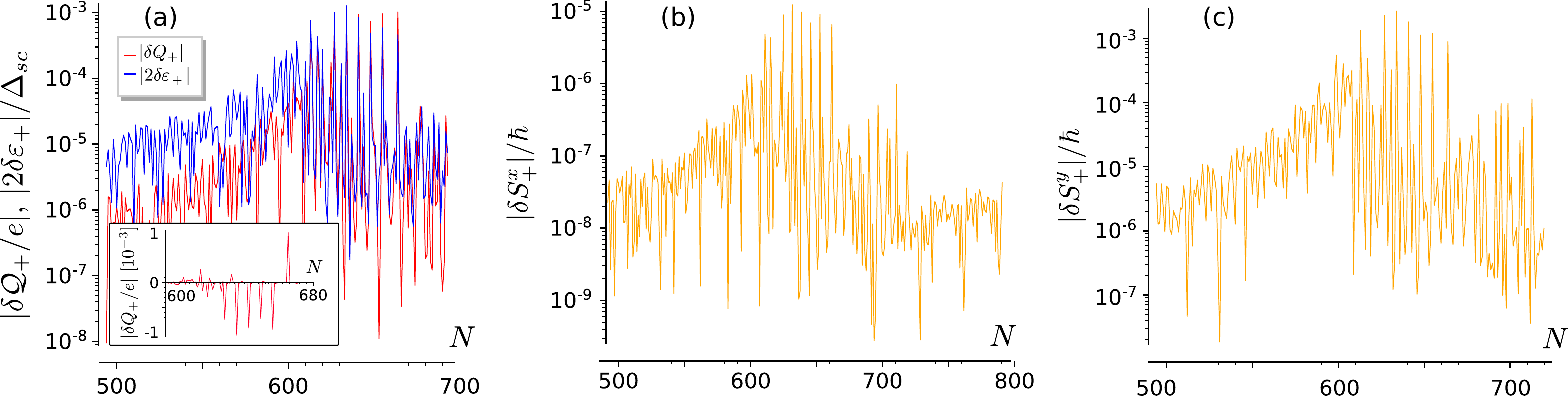}
	\caption{ (a) Charge $\delta Q_+$  as well as components of the spin (b) $\delta S^x_+$ along the $x$ and (c) $\delta S^y_+$ along $y$ axis of the MBSs induced by electron-phonon interaction as function of number of sites $N$ found numerically in the spinful  Rashba lattice model. The $z$ component of the spin, $S^z_{\pm}$, remains zero due to a symmetry of the system~\cite{Szumniak2017}. A finite energy splitting $|2\delta \varepsilon_+|$ (blue solid line [see also Fig.~\ref{fig:Rashba1_ab} (a)]) gives rise to a finite charge and spin of the MBSs. Both charge and spin oscillate with increasing the system length $N$, showing resonant enhancement similar to the energy splitting [cf. Fig.~\ref{fig:VHS_vs_L} and Fig.~\ref{fig:Rashba1_ab} (a)]. The parameters chosen are $(E_{so}, \Delta_{sc}, \Delta_Z, \Omega_1) = (0.6, 0.2, 0.4, 0.22)\; \mathrm{meV}$, corresponding to $t=10\mathrm{meV}$, coherence length $\xi = 50a_0$. The rest of the parameters are the same as for Fig.~\ref{fig:VHS_vs_L}. }
	\label{fig:charge}
\end{figure*}

\begin{figure*}
	\includegraphics[width=0.95\linewidth]{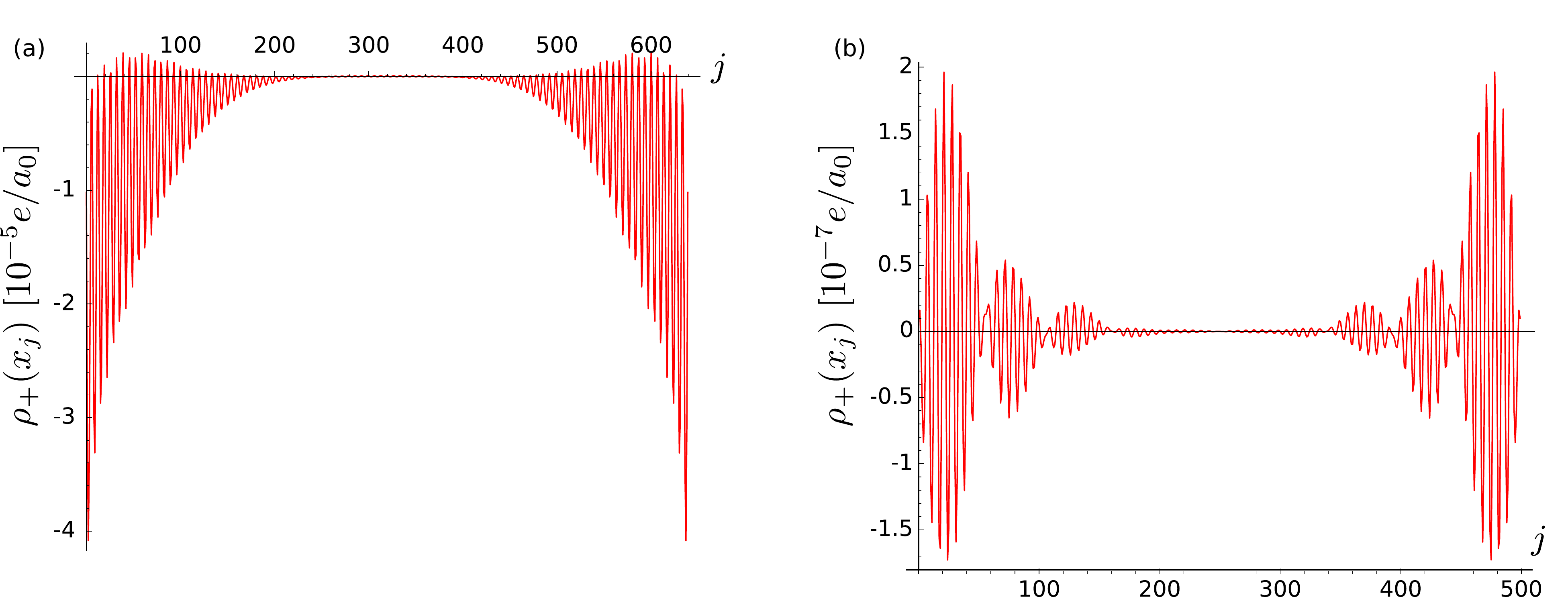}
	\caption{Charge density $\delta\rho_+(x)$ of the MBSs induced by the electron-phonon interaction as function of lattice site $j$ (a) close to the resonance, $N=640$  [see Fig.~\ref{fig:charge} (a)], and (b) away from the resonance, $N=500$. While in the non-resonant case (b), the charge density changes its sign and oscillates on the characteristic scale $c_s/\Delta$ corresponding to the phonon wavelength, in case of the resonance (a), the total charge on the negatively charged sites (for the given values of parameters) is larger than the one on the positively charged site such that the overall charge carried by the MBSs is non-zero. The parameters chosen are the same as for Fig.~\ref{fig:charge}.}
		\label{fig: density}
	\end{figure*}

In the absence of electron-phonon interactions, the energies of the MBSs are exponentially small in system size. Moreover, particle-hole symmetry implies that both MBSs carry zero charge (up to exponentially small corrections). This is one of the attractive features of MBSs for the purpose of qubits since this keeps them insensitive to any fluctuations of effective electromagnetic fields in the surroundings.
However, in the presence of electron-phonon interactions a finite energy splitting $|2\delta \varepsilon_+|$ gives rise to a finite charge of the MBSs which is still less than $e$ since each MBS is a superposition of electron and hole states. In this subsection we calculate the charge density $\delta \rho_\pm(x) =  e\langle \pm|\Psi^\dag(x) \Psi(x) |\pm\rangle$ and the overall charge $\delta \mathcal{Q}_\pm = \int dx\; \delta \rho_{\pm}(x)$  of the MBSs induced by the electron-phonon interaction in the presence of VHS in the phonon density of states. We note that electrons and holes contribute to the charge with opposite signs, which is taken into account here since the operator $\Psi(x)$ is a superposition of electron and hole quasiparticle operators [see Eq.~(\ref{eqn:Heisenberg1})]. The effective Hamiltonian, in which the phonon degrees of freedom are integrated out, has the following form:
\begin{align}
H_{\mathrm{eff}} = \sum\limits_{\alpha} (\varepsilon_\alpha + \delta \varepsilon_\alpha) \tilde{c}_\alpha^\dag \tilde{c}_\alpha,
\end{align}
where in the leading order of perturbation theory $\delta\varepsilon_\alpha \approx \mathrm{Re}\; \Sigma^R_{\alpha \alpha}(\varepsilon_\alpha)$, and fermionic operators $\tilde{c}_\alpha(t)$ in Heisenberg representation can be related to the unperturbed quasiparticle operators $c_\alpha(t)$ as\cite{Kamenev09}:
\begin{align}
\begin{split}
c_\alpha(t) &=\; \tilde{c}_\alpha(t) \\&+\; \sum\limits_{\beta\neq \alpha}\int dt' dt'' G^R_\alpha(t-t')\Sigma^R_{\alpha\beta}(t'-t'')\tilde{c}_\beta(t''),
\end{split}
\end{align}
which is convenient to rewrite in energy representation as:
\begin{align}
c_\alpha(\varepsilon) &=\; \tilde{c}_\alpha(\varepsilon) + \frac{1}{\varepsilon-\varepsilon_\alpha + i0^+} \sum\limits_{\beta\neq \alpha} \Sigma^R_{\alpha \beta}(\varepsilon)\tilde{c}_\beta(\varepsilon).
\label{eqn:c-tilde}
\end{align}
Here, $c_\alpha(\varepsilon) = \int dt\; c_\alpha(t) e^{i\varepsilon t}$, $\tilde{c}_\alpha(\varepsilon) = \int dt\; \tilde{c}_\alpha(t) e^{i\varepsilon t}$ is the Fourier transform of the operator in Heisenberg representation. Combining Eq.~(\ref{eqn:c-tilde}) with Eqs.~(\ref{eqn:Heisenberg1})--(\ref{eqn:Heisenberg2}), we obtain the spinors corresponding to the perturbed fermion states $\tilde{\Phi}_\alpha = \Phi_{\alpha} + \delta \Phi_{\alpha}$:
\begin{align}
\delta \Phi_\alpha = \sum\limits_{ \beta \neq \alpha} \Phi_\beta \frac{\Sigma^R_{\beta \alpha}(\varepsilon_\alpha)}{\varepsilon_\alpha - \varepsilon_\beta}.
\label{eqn:delta-Phi}
\end{align}
Thus, the electron-phonon interaction hybridizes the MBSs with bulk quasiparticle states. The charge density carried by the perturbed fermionic subgap states corresponding to $\tilde{c}_{\pm}$ can be calculated as  
\begin{align}
&\delta \mathcal{\rho}_\pm(x) =\; e\sum\limits_{mn}\int dx_1dx_2\;\label{eqn:charge}\\
&\hspace{30pt} \times \frac{\rho_{\pm m}(x)\rho_{mn}(x_1)\mathrm{Re}\left\{W^R_{n}(x_1-x_2)\right\} \rho_{n\pm}(x_2)}{\varepsilon_\pm - \varepsilon_m}. \nonumber
\end{align}
We note that particle-hole symmetry implies $\delta\rho_+(x) =-\delta\rho_-(x)$. 
For the spinful model considered in Sec.~\ref{sec:Rashba}, the charge and spin of the MBSs induced by electron-phonon interaction can be calculated similarly to Eq.~(\ref{eqn:charge}):
\begin{align}
&\delta \rho_\pm(x_j) = e\sum\limits_{mn}\sum\limits_{j_1,j_2}a_0^2  \label{eqn:spin-Rashba}
\\&\times \frac{\rho_{\pm m}(x_j)\rho_{mn}(x_{j_1})\mathrm{Re}\left\{W^R_{n}(x_{j_1}-x_{j_2})\right\} \rho_{n\pm}(x_{j_2})}{\varepsilon_\pm - \varepsilon_m}, \nonumber
\\
& \delta \mathcal{Q}_\pm = \sum\limits_j \rho_\pm(x_j),\\
&\delta S^{k}_\pm =\sum\limits_{mn}\sum\limits_{j_1,j_2,j_3}a_0^3\;\\&\times\; \frac{s^{k}_{\pm m}(x_{j_1})\rho_{mn}(x_{j_2})\mathrm{Re}\left\{W^R_{n}(x_{j_2}-x_{j_3})\right\} \rho_{n\pm}(x_{j_3})}{\varepsilon_\pm - \varepsilon_m},\nonumber
\end{align}
where $s^{k}_{\pm m}(x_j)= (\hbar/2) \Phi^\dag(x_j) \sigma^{k} \tau_z \Phi(x_j)$,
with $k=x,y,z$.

 We calculate the overall charge $\delta \mathcal{Q}_\pm$ induced by the electron-phonon interaction numerically using the spinful lattice Rashba model. The total MBS charge  $\delta \mathcal{Q}_\pm$ as a function of number of sites $N$  is shown in Fig.~\ref{fig:charge}(a).  The charge density $\delta\rho_+(x)$ as function of position is shown in Fig.~\ref{fig: density}. Similarly, we calculate numerically the spin of the perturbed fermionic states $|\pm\rangle$ for the spinful lattice Rashba model [see Fig.~\ref{fig:charge}(b, c)] using Eq.~(\ref{eqn:spin-Rashba}). Thus, electron-phonon interaction induces both a finite charge $\delta Q_{\pm}$ and spin $\delta S^{x,y}_\pm$ for the spinful model. We note that the $z$ component of the spin, $\delta S^z_\pm$ (along the direction set by the Rashba spin-orbit interaction), remains zero due to the symmetry of the system\cite{Szumniak2017}. Both charge and spin oscillate with increasing system size showing resonant enhancement similar to the energy splitting.
It is important to note that non-zero charge and spin induced by the electron-phonon interaction makes it possible to detect the splitting via charge and spin measurements. On the other hand, the finite charge and spin couple the hybrized MBSs to external electrical and magnetic noise sources which will affect the decoherence properties of topological qubits formed from such MBSs.

	\section{Conclusions \label{sec:conclusions}}
	
	In this work we studied the effect of electron-phonon interactions on MBSs in topological nanowires. We have shown that at zero temperature such perturbations do not lift the ground state degeneracy and do not induce a finite level broadening. However, at finite temperatures absorption of thermal phonons makes it possible to promote the electron system from the ground state to a delocalized state with energy above the quasiparticle gap $\Delta$. The level broadening (inverse lifetime) decays exponentially with the inverse of the temperature. This source of decay is an intrinsic many-body effect rather than due to the presence of an externally coupled environment. The lifetime the system stays in the ground state is estimated to be of order $10\;\mathrm{\mu s}$ for temperatures of order $0.1\; \mathrm{K}$ and quasiparticle gap of order $1\;\mathrm{K}$. Furthermore, the coherent absorption/emission of phonons at the ends of the topological nanowire results in lifting of the degeneracy of the ground state at finite temperatures. The resulting energy splitting between MBSs decays as a power-law rather than exponentially with increasing system size. However, experimental observation of this splitting can be complicated since it remains less than the level broadening of the MBSs.  As a consequence of the splitting induced by the phonons, the MBSs acquire a finite charge and spin  which we calculated as a function of position as well as of the system size. This  opens up the possibility to detect the splitting via charge and spin measurements. On the other hand, it also exposes the MBSs to external electrical and magnetic noise sources that will affect the decoherence properties of topological qubits formed from such MBSs. It will be interesting to investigate this problem further in future work.

	We also found that if the motion of phonons is quantized in the transverse direction of the nanowire, the presence of Van Hove singularities at the bottom of the phonon modes enhances the absorption of phonons if the singularity energy is close to the quasiparticle gap. In this case the lifetime is estimated to be less than nanoseconds for the same values of temperature and the gap as above. The energy splitting and broadening of the MBSs show resonant peaks at  energies where a VHS  coincides with the energy of a bulk quasiparticle state. Close to the resonance, the energy splitting of the MBSs becomes comparable to the broadening (or even larger) and, therefore, can be observed experimentally. The results obtained analytically for the low-energy continuous model are also confirmed numerically for a spinful lattice Rashba model. 
	
	Finally, we remark that the pronounced resonances for level splitting and broadening found in this work are a unique manifestation of the presence of MBSs and thus can serve as further experimental signatures  in the search of  MBSs in topological nanowires.

	\section*{Acknowledgments}
	We would like to thank Chen Hsu and Dmitry Miserev for useful discussions. This work was supported by the Swiss National Science Foundation (Switzerland) and by the NCCR QSIT. This project received funding from the European Union's Horizon 2020 research and innovation program (ERC Starting Grant, grant agreement No 757725) 
	and JSPS Kakenhi Grant No. 16H02204.
	\appendix
	
	\section{Solutions of BdG equations \label{app:BdG}}
In this section, we solve the BdG equation	[see Eq.~(\ref{eqn:BdG})] subjected to boundary conditions (BCs), see Eqs.~(\ref{eqn:linearized-BCs1})--(\ref{eqn:linearized-BCs2}).
	In order to simplify the calculations we exploit the fact that the system has inversion symmetry. More specifically,
	the BdG equation
	and BCs
	are invariant under the symmetry operator $\hat{I}$ that acts on the spinors $\Phi$ in the following way:
	\begin{align}
	&\hat{I}
	\begin{pmatrix}
	\phi_+(x)\\
	\phi_-(x)\\
	\bar{\phi}_+(x)\\
	\bar{\phi}_-(x)
	\end{pmatrix}
	=
	\begin{pmatrix}
	\phi_-(L-x)e^{-ik_FL}\\
	\phi_+(L-x)e^{ik_F L}\\
	-\bar{\phi}_-(L-x)e^{ik_FL}\\
	-\bar{\phi}_+(L-x)e^{-ik_FL}\, .
	\end{pmatrix}
	\label{def:I}
	\end{align} 
	Since $\hat{I}^2 = \mathbb{1}$, the solutions of the BdG equations are either even or odd 
	under inversion.
	\subsection{Bound states}	
	 A general solution of Eq.~(\ref{eqn:BdG}) with energy  below the gap, $|\varepsilon| < \Delta$, has the form
	\begin{align}
	\begin{split}
	\Phi(x) = A\left(\Phi_{+,-} e^{-\kappa x} \pm \Phi_{-,-} e^{ik_FL - \kappa(L-x)} \right) \\
	+ B\left(\Phi_{-,+} e^{-\kappa x} \pm \Phi_{+,+} e^{-ik_FL - \kappa(L-x)} \right),
	\label{eqn:solution}
	\end{split}
	\end{align}
	where the upper (lower) sign corresponds to an (even) odd solution with respect to $\hat{I}$, $\Phi_{+,\pm} = \begin{pmatrix}-i,&0,&0,&(\varepsilon \pm iv_F \kappa)/\Delta\end{pmatrix}^T$,
	$\Phi_{-,\pm} = \begin{pmatrix}0,&-i,&-(\varepsilon \pm iv_F \kappa)/\Delta,&0\end{pmatrix}^T$, and the decay parameter $\kappa = \sqrt{\Delta^2 - \varepsilon^2}/v_F$, $A$ and $B$ are coefficients which are chosen to satisfy the BCs. The BCs [see Eq.~(\ref{eqn:linearized-BCs1})--(\ref{eqn:linearized-BCs2})] result in the following equations for $A$ and $B$:
		\label{def:A}\label{def:B}\label{def:Phi_pmpm}
	\begin{align}
	&A\left(1 \pm e^{ik_F L} e^{-\kappa L} \right) + B\left(1 \pm e^{-ik_F L } e^{-\kappa L} \right) = 0,	\label{def:system-MBS1}\\
	\begin{split}
	&A\left(\varepsilon - i\kappa v_F\right)\left( 1\mp e^{ik_FL e^{-\kappa L}}\right) +
	B\left(\varepsilon + i\kappa v_F \right)\\&\times\left( -1 \pm e^{-ik_FL}e^{-\kappa L} \right) =0.
	\end{split}
	\label{def:system-MBS2}
	\end{align}
	Solving the linear system defined by Eqs.~(\ref{def:system-MBS1})--(\ref{def:system-MBS2}), we obtain that an even (odd) solution with energy below the gap, $|\varepsilon| < \Delta$, is of the form $\Phi_{\pm}(x) = \left[\Phi_L(x) \pm i\Phi_R(x)\right]/\sqrt{2}$ , where
	\begin{align}
	\Phi_L(x) &=\; 
	\frac{1}{\sqrt{\xi}}
	\begin{pmatrix}
	-i\\
	i\\
	i\\
	-i
	\end{pmatrix} e^{-x/\xi}\, ,\\
	\Phi_R(x) &=\;
	\frac{1}{\sqrt{\xi}}
	\begin{pmatrix}
	e^{-ik_F L}\\
	-e^{ik_F L}\\
	e^{ik_F L}\\
	-e^{-ik_FL}
	\end{pmatrix}e^{-\left(L-x\right)/\xi},
	\end{align}	
	and the energies satisfy Eq.~(\ref{eqn:MBS-energy}). 
	\subsection{Bulk states}
	 A general solution of Eq.~(\ref{eqn:BdG}) with energy above the gap, $|\varepsilon_n| > \Delta$, has the form:
	 \begin{align}
	& \Phi_n(x) = A_n\left[\Phi_{n,+, -}e^{ik_nx} \pm \Phi_{n,-,-} e^{ik_F L + ik_n(L-x)} \right] \nonumber
	\\
	 &+ B_n\left[\Phi_{n,-,+} e^{ik_nx} \pm \Phi_{n,+,+} e^{-ik_FL} e^{ik_n(L-x)} \right],
	 \end{align}
	 where the upper (lower) sign corresponds to an even (odd) solution with respect to $\hat{I}$, 
	 $\Phi_{n,+,\pm} = 
	 \begin{pmatrix}
	 -i,&
	 0,&
	 0,&
	 (\varepsilon_n \pm \zeta_n)/\Delta
	 \end{pmatrix}^T$,
	 $\Phi_{n,-,\pm} = 
	 \begin{pmatrix}
	 0,&
	 -i,&
	 -(\varepsilon_n \pm \zeta_n)/\Delta,&
	 0
	 \end{pmatrix}^T$,
	 $\zeta_n = \sqrt{\varepsilon_n^2 - \Delta^2}$, and $k_n = \zeta_n/v_F$.
	 The BCs defined by ~Eqs.~(\ref{eqn:linearized-BCs1})--(\ref{eqn:linearized-BCs2}) result in the following equations for $A_n$, $B_n$:
	 \begin{align}
	 &A_n\left(1 \pm e^{i(k_F + k_n)L}\right) = - B_n \left(1 \pm e^{-i(k_F - k_n)L }\right) \label{eqn:bulk1},\\
	 \begin{split}
	 &A_n\frac{\varepsilon_n - \zeta_n}{\varepsilon_n + \zeta_n}\frac{1\mp e^{i(k_F+k_n)L}}{ 1\mp e^{-i(k_F-k_n)L}}=
	 B_n.
	 \end{split}
	 \label{eqn:bulk2}
	 \end{align}
	Solving this linear system of equations, we obtain that the
	energies of the bulk eigenstates satisfy the following condition:
	\begin{align}
	\sin \left(k_n L\right) = \pm\frac{\zeta_n}{\varepsilon_n}\sin\left(k_F L\right),
	\end{align}
	where the upper (lower) sign corresponds to even (odd) solutions with respect to $\hat{I}$, which themselves are of the form:
	\begin{align}
	&\Phi_n = \frac{1}{\mathcal{N}_n\Delta\sqrt{8L} }\left[
	\begin{pmatrix}
	-i\Delta\left(1\pm e^{i(k_n-k_F)L}
	\right)
	\\
	i\Delta\left(1\pm  e^{i(k_F+k_n)L}\right)\\
	(\varepsilon_n+\zeta_n)
	\left(1\pm e^{i(k_F+k_n)L}\right)\\
	(\varepsilon_n-\zeta_n)\left(1\pm e^{i(k_n-k_F)L}
	\right)
	\end{pmatrix}
	e^{ik_nx}
	\right.\\&+\left.
	\begin{pmatrix}
	i\Delta\left(\pm e^{ik_F L}+e^{ik_nL}\right)\\
	-i\Delta\left(\pm e^{ik_F L}+e^{ik_nL}
	\right)\\
	-(\varepsilon_n-\zeta_n)\left(\pm e^{ik_F L}+e^{ik_nL}
	\right)
	\\
	-(\varepsilon_n+\zeta_n)
	\left(\pm e^{-ik_Fl}+e^{ik_nL}\right)
	\end{pmatrix}
	e^{ik_n(L-x)}
	\right],\nonumber
	\end{align}
	where the normalization factor is given by
	\begin{align}
	\mathcal{N}_n = 
	\sqrt{\frac{\varepsilon_n^2}{\Delta^2}
		\left[1 \pm \cos \left(k_F L\right)\cos\left(k_n L\right)\right] - \frac{\zeta_n^2}{\Delta^2}\sin^2\left( k_F L\right)}.
	\end{align}
	
	\section{Overlap integrals \label{app:matrix}}
	The matrix elements $P_{L,n} = 	\int \rho_{L,n}(x) dx$ can be calculated straightforwardly:
	\begin{multline}
 P_{L,n} =
	 \frac{\sqrt{8} v_F 	\left[ \zeta \cos\left(k_n L
	 	\right) \pm \zeta\cos \left(k_F L\right) -\Delta\sin \left(k_n L \right)  \right]}{\mathcal{N}_n \varepsilon_n^2\sqrt{\xi L}}\\\times e^{i\left(k_n L + \frac{\pi}{2}\right)},
\end{multline}
where the upper (lower) sign corresponds to even (odd) bulk modes $n$  
with respect to the inversion symmetry $\hat{I}$.
The overlap integrals containing exponentials from the phonon fields can be similarly found as
\begin{align}
\int dx\; \rho_{L,n}(x) e^{iqx} = P_{L,n} \frac{\varepsilon_n^2\left[1 - iqv_F/(2\Delta) \right]}{\varepsilon_n^2 - q^2 v_F^2 - i qv_F \Delta}.
\end{align}
We note that inversion symmetry implies that $P_{n,R} = \pm P_{L,n}$ and $\int dx\; i\rho_{n,R}(x) e^{iq(L-x)} = \pm\int dx\; \rho_{L,n}(x) e^{iqx}$. Therefore, we calculate
\begin{align}
&\int dxdx'\; i\rho_{L,n}(x) \sin \left[q(x'-x)\right] \rho_{n,R}(x') = \nonumber 
\\& \pm \mathrm{Im}\;\left\{ \left[\int dx \rho_{L,n}(x)e^{-iqx}\right]^2 e^{-iqL}  \right\} = \nonumber \\
&  \pm |P_{L,n}|^2 \varepsilon_n^4\; \mathrm{Im}\left\{\left[ \frac{1 + iq v_F/(2\Delta)}{\varepsilon_n^2 - q^2 v_F^2 - iqv_F \Delta} \right]^2 e^{-iqL}\right\},
\label{eqn:LR_sin}
\end{align}	
where the upper (lower) sign corresponds to the even (odd) bulk modes $n$. 
For $q \sim \Delta/c_s \gg \varepsilon_n/v_F$, we obtain a simple approximate expression,
\begin{multline}
\int dxdx'\; \rho_{L,n}(x) \sin \left[q(x'-x)\right] \rho_{n,R}(x') \\\approx \pm|P_{L,n}|^2
\frac{\varepsilon_n^4}{4\Delta^2 q^2 v_F^2 }\sin \left( qL \right).
\label{eqn:LL_sin}
\end{multline}
Similarly, we calculate
\begin{multline}
\int dxdx' \rho_{L,n}(x)\cos[q(x'-x)]\rho_{n,L}(x') \\= |P_{L,n}|^2 \varepsilon_n^4 \frac{1+q^2v_F^2/(4\Delta^2)}{(\varepsilon_n^2 -q^2 v_F^2)^2 + q^2v_F^2\Delta^2}.
\label{eqn:LL_cos}
\end{multline}	
For $q \sim \Delta/c_s \gg \varepsilon_n/v_F$, the expression given by ~Eq.~(\ref{eqn:LL_cos}) simplifies to
\begin{align}
&\int dxdx' \rho_{L,n}(x)\cos[q(x'-x)]\rho_{n,L}(x') \\
& \hspace{130pt} \approx |P_{L,n}|^2 \frac{\varepsilon_n^4}{4\Delta^2 q^2 v_F^2}. \nonumber
\end{align}
We also calculate the overlap integrals containing decaying exponentials from the phonon fields:
\begin{align}
&\int dx\; \rho_{L,n}(x) e^{-Qx} = P_{L,n} \frac{\varepsilon_n^2\left[1 + Qv_F/(2\Delta) \right]}{\varepsilon_n^2 + Q^2 v_F^2 + Qv_F \Delta},\\
\begin{split}
&\int dx dx' \; i\rho_{L,n}(x)\rho_{n,R}(x') e^{-Q|x-x'|} =\\ &\pm e^{-QL} |P_{L,n}|^2 \varepsilon_n^4\left[\frac{1- Qv_F /(2\Delta)}{\varepsilon_n^2 + Q^2 v_F^2 - Qv_F \Delta}\right]^2.
\end{split}
\end{align}
Again, the upper (lower) signs correspond to  even (odd) bulk modes $n$.

\section{Effective interaction \label{app:eff-interaction}}
The effective interaction $W^R_n(x,\varepsilon)$ is defined in Eq.~(\ref{eqn:effective-interaction}), where, again, we neglect boundary effects for the phonons and assume that  the phonon system is translationally invariant (in contrast to the TSC).
Integrating over the phonon frequency $\omega$ we obtain $W^R_n(x, \varepsilon) = \int W^R_n(q, \varepsilon) e^{iqx} dq/(2\pi)$ for 1D phonons in infinite space, where the Fourier transform $W_n(q, \varepsilon)$ is given by
\begin{align}
\begin{split}
&W^R_n(q,\varepsilon) = 
\frac{1}{2} \frac{\Omega_q(\varepsilon-\varepsilon_n)}{(\varepsilon-\varepsilon_n + i0^+)^2 - \Omega_q^2} \coth \left( \frac{\Omega_q}{2T} \right)\\
&\hspace{30pt}+ \frac{1}{2} \frac{\Omega_q^2}{(\varepsilon-\varepsilon_n + i0^+)^2 - \Omega_q^2}\tanh \left( \frac{\varepsilon_n}{2T} \right).
\end{split}
\label{eqn:W_q}
\end{align}
In the following, we take $\varepsilon=0$.
In case of a linear phonon spectrum, $\Omega_q = c_s q$, the integration over $q$ can be performed straightforwardly using contour integration: the expression in Eq.~(\ref{eqn:W_q}) has the following poles in the upper half plane: $q_0 = -\varepsilon_n/c_s + i0^+$ and $q_k = 2\pi i k T$, where $k\in\mathbb{N}$. The pole $q_0$ vanishes at $T=0$, and the contribution from this pole, $W^R_{a,n}$, given by Eq.~(\ref{eqn:W-absorption}), corresponds to absorption of a thermal phonon with energy $\varepsilon_n$ and momentum $q_0$. The remaining poles $q_k$ give the contribution $W^R_{v,n}$, see Eq.~(\ref{eqn:W-virtual}), describing the effective interaction due to exchange of virtual phonons. In the limit $T=0$, one can estimate this contribution as
\begin{align}
&W^R_{v,n}(x, \varepsilon = 0) \approx\; \gOne^2\sum\limits_{k=1}^{\infty}\frac{ T}{2\varepsilon_n} \frac{2\pi k T}{c_s}\exp\left\{-\frac{2\pi k T |x|}{c_s} \right\}\nonumber  \\
&\hspace{130pt}= \frac{\gOne^2 c_s}{4\pi \varepsilon_n x^2}.
\end{align}
We note that $W^R_{v,n}$ results from the integration of the first term in~Eq.~(\ref{eqn:W_q}) containing the phonon distribution function, and, hence, the exponential decay is determined by the temperature of the phonon bath. The contribution $W^R_{a,n}$ given by Eq.~(\ref{eqn:W-absorption}) depends both on electron and phonon distributions.

In the case when the spectrum $\Omega_q^2 = \Omega_1^2 + c_s^2q^2$ describes a higher phonon mode, the expressions $\Omega_q  \coth (\Omega_q/2T)$ and $\Omega_q^2$ are still single-valued analytic functions of $q$, and the absorption of thermal phonons is described by the contribution from the pole $q = -\sqrt{(\varepsilon_n-i0^+)^2 - \Omega^2}/c_s$ for $\varepsilon_n > \Omega_1$ and $q = i\sqrt{\Omega_1^2 - \varepsilon_n^2}/c_s$ for $|\varepsilon_n| < \Omega_1$, resulting in the expression  $W^R_{a,n}$ given in Eq.~(\ref{eqn:WR_VHS}).

\begin{figure}[t]
	\vspace{1em}
	\includegraphics[width=0.6\linewidth]{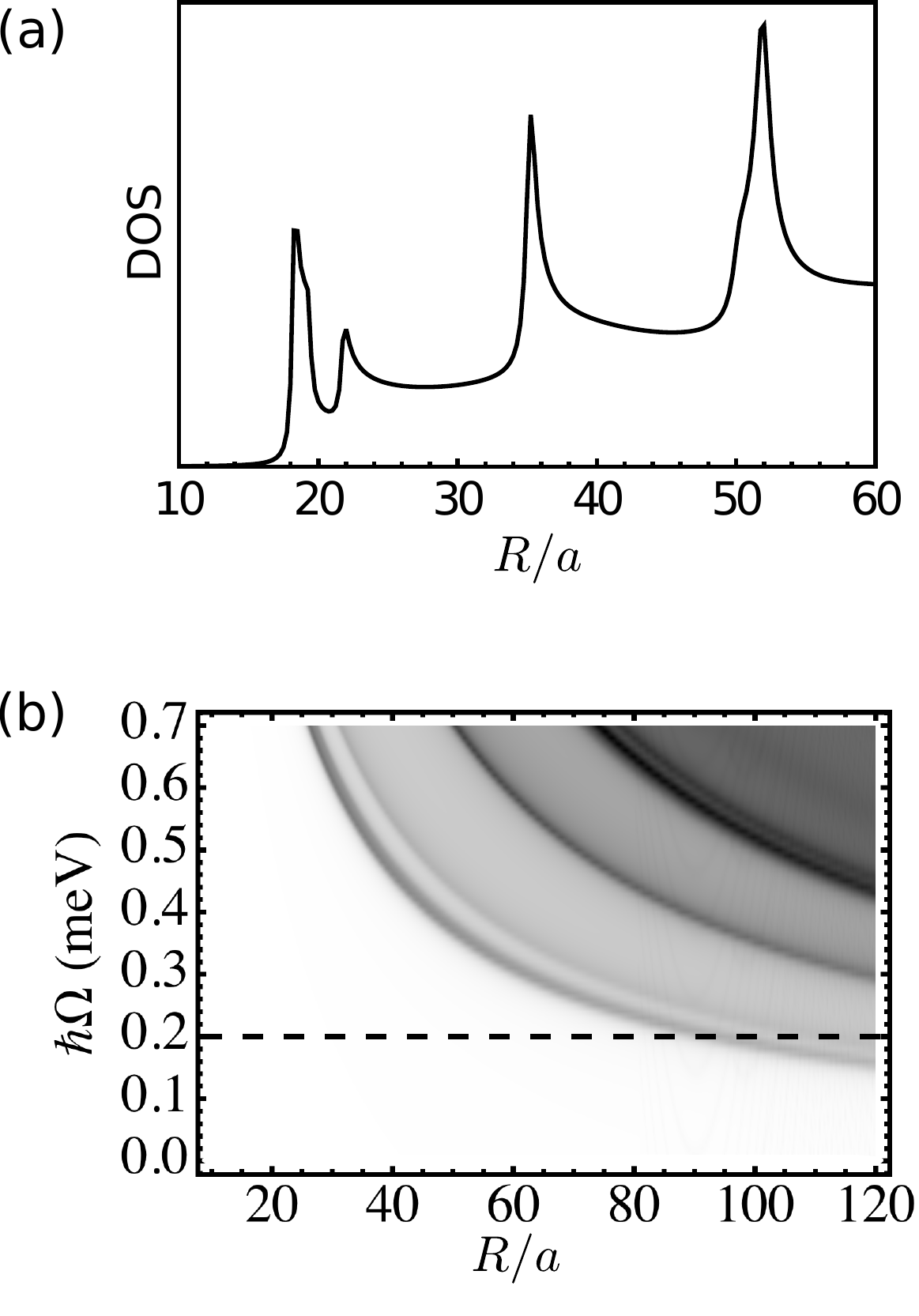}
	\caption{Phonon density of states (DOS) corresponding to the phonon dispersion in Eq.~(\ref{eq:PC}) with $c_{t}/c_{l}=0.5$ at fixed energy $\Delta=1\;\mathrm{meV}$  (a) as a function of the nanowire radius $R$ and (b) as function the phonon energy $\hbar \Omega$ and $R$.
		The pronounced peaks correspond to VHSs, which have energies inversely proportional to $R$.	The bottom of phonon branch $\hbar\Omega = 0.2\;\mathrm{meV}$ (dashed horizontal line) corresponds to $R = 100a = 50\;\mathrm{nm}$. }
	\label{fig:NW}
\end{figure}

\section{Van Hove singularity in nanowires \label{app:VHS}}

In order to evaluate the effect of VHSs in a realistic system, we consider longitudinal phonons of a homogeneous nanowire of radius $R$ with clamped surface boundary conditions\cite{Kloeffel2014,Trif2008,Cleland2002}.
In the continuum limit, the dispersion relation for nontorsional longitudinal phonons is given by the Pochhammer-Chree frequency equation, which reads
\begin{gather}
(b_{t}^2-q^2)^2		J_0(b_{l} R) J_1(b_{t} R)
+ 4q^2 b_{l} b_{t}		J_1(b_{l} R) J_0(b_{t} R)
\nonumber\\=
2 (b_{l}/R) (b_{t}^2+q^2)	J_1(b_{l} R) J_1(b_{t} R),
\label{eq:PC}
\end{gather}
where $J_n(x)$ are the Bessel functions of the first kind, and  $b_{l}^2=\Omega_q^2/c_{s}^2-q^2$, $b_{t}^2=\Omega_q^2/c_{t}^2-q^2$ with $c_{s}^2 = \lambda + 2\mu$ and $c_{t}^2 = \mu$ the longitudinal and shear speed of sound, respectively, given in terms of the Lam\'{e} parameters $\lambda$ and $\mu$. Generally, Eq.~(\ref{eq:PC}) is a good approximation for longitudinal phonon modes in a nanowire at small momenta.
For each momentum $q$, the above equation is satisfied by multiple values of the frequency $\Omega_q$.
Therefore, the phonon dispersion exhibits multiple branches $\Omega_{q,j}$:
This is a physical consequence of the confinement of the phonon modes in the  transverse direction.
The average energy spacing between the phonon branches at $q=0$ is of the order of $\hbar c_{l}/R$.
We consider the phonon dispersion at small momenta calculated numerically from Eq.~(\ref{eq:PC}) using $c_{t}/c_{l}=0.5$ as a function of $R/a$.
The resulting phonon dispersion exhibits multiple VHSs at energies $\Omega_j$, as one can see from the peaks in the phonon density of states in Fig.~\ref{fig:NW}: 
	the energies $\Omega_j$ decrease and the spacing between them shrinks as the the wire radius increases.

\section{Splitting and broadening in the presence of VHS \label{app:estimations}}
\subsection{Broadening}
We rewrite Eq.~(\ref{eqn:VHS-Gamma}) using expressions for overlap integrals we found earlier, Eq.~(\ref{eqn:LL_cos}),
\begin{multline}
\gamma_L = \frac{g_1^2}{4c_s}\sum\limits_n |P_{Ln}|^2 \frac{\varepsilon_n^6}{\sinh(\varepsilon_n/T) \sqrt{\varepsilon_n^2 - \Omega_1^2}}\\\times \frac{1+Q_n^2 v_F^2/(4\Delta^2)}{(\varepsilon_n^2 - Q^2_n v_F^2)^2 + Q^2_nv_F^2 \Delta^2},
\label{eqn:gamma_L}
\end{multline}
where $Q_n = \sqrt{\varepsilon_n^2 - \Omega_1^2}/c_s$. In the limit of long nanowire $L/\xi \to \infty$ one can estimate contribution from the non-resonant modes by replacing the sum over $n$ by an integral, and then changing the integration variable from $n$ to $\varepsilon_n\equiv \varepsilon$,
\begin{align}
dn &=\; \left|\frac{d \varepsilon_n}{dn}\right|^{-1} d\varepsilon \approx \frac{L}{\pi v_F} \frac{\varepsilon\, d \varepsilon}{\sqrt{\varepsilon^2 - \Delta^2}}\,.
\end{align}
Thus, since  the matrix element $|P_{L,n}|^2\sim \xi/L$ for $\varepsilon_n \sim \Omega_1 \approx \Delta$, we can estimate:
\begin{align}
\begin{split}
\gamma_L &\sim\; \frac{g_1^2 v_F}{c_s \Delta L}\int\limits_{\Omega_1}^{+\infty} \frac{L\varepsilon d\varepsilon}{\pi  v_F\sqrt{\varepsilon^2 - \Delta^2}} 
\frac{e^{-\varepsilon/T}  \varepsilon_n^6}{\sinh(\varepsilon_n/T) \sqrt{\varepsilon^2 - \Omega_1^2}}\\&\times \frac{1+(\varepsilon^2 - \Omega_1^2) (v_F/c_s)^2/(4\Delta^2)}{[\varepsilon_n^2 - (\varepsilon^2 - \Omega_1^2) (v_F/c_s)^2]^2 + (\varepsilon^2- \Omega_1^2)(v_F/c_s)^2 \Delta^2}\\
&\approx \frac{g_1^2}{\pi c_s}\frac{\Delta^2}{\sqrt{\Omega_1^2 - \Delta^2}}e^{-\Omega_1/T}\ln \left(\frac{T}{\Gamma}\right). 
\end{split}
\end{align}
\subsection{Energy splitting}
Here we show that the energy splitting  $\delta \varepsilon_{+}^>$ remains smaller than the broadening $\gamma_{L(R)}$. Using Eq.~(\ref{eqn:LL_sin}), the contribution $\delta\varepsilon_{+,>}$ can be written as
\begin{align}
&\delta \varepsilon_{+}^> = \frac{g_1^2}{4c_s}\sum\limits_{n,\varepsilon_n > 
\Omega_1} \frac{(-1)^n |P_{L,n}|^2 \varepsilon_n^6}{\sinh(\varepsilon_n/T)\sqrt{\varepsilon_n^2 - \Omega_1^2}}\nonumber\\ 
&\hspace{50pt}\times \mathrm{Im}\left\{\left[\frac{1 + iQ_n v_F/(2\Delta)}{\varepsilon_n^2 - Q_n^2 v_F^2 - iQ_nv_F \Delta} \right]^2 e^{iQ_nL}\right\} \nonumber \\
&=\; \frac{g_1^2}{4c_s}\sum\limits_{n,\varepsilon_n > 
\Omega_1} |P_{Ln}|^2 \frac{\varepsilon_n^6}{\sinh(\varepsilon_n/T) \sqrt{\varepsilon_n^2 - \Omega_1^2}}\nonumber\\
&\hspace{50pt}\times \frac{1+Q_n^2 v_F^2/(4\Delta^2)}{(\varepsilon_n^2 - Q^2_n v_F^2)^2 + Q^2_nv_F^2 \Delta^2}\sin \chi_n,
\label{eqn:dE_gtr}
\end{align}
where the phase $\chi_n$ is defined as
\begin{align}
\begin{split}
\chi_n &=\; \pi n +Q_nL + 2\arctan\left(\frac{Q_n v_F}{2\Delta} \right) \\
&\hspace{70pt}- 2\arctan\left(\frac{Q_nv_F \Delta}{\varepsilon_n^2 - Q_n^2 v_F^2}\right).
\end{split}
\end{align}
Comparing Eq.~(\ref{eqn:dE_gtr}) with Eq.~(\ref{eqn:gamma_L}), one can see that $|\delta \varepsilon_+|\leq \gamma_{L,R}$.


\begin{thebibliography}{99}
	\bibitem []{Kitaev2001}%
	{A.~Y.~Kitaev, \href{\doibase 10.1070/1063-7869/44/10S/S29}{Physics-Uspekhi \textbf{44}, 131 (2001)}.}
	%
	\bibitem []{Kitaev2003}%
	{A.~Kitaev, \href{\doibase 10.1016/S0003-4916(02)00018-0} {Ann. Phys. \textbf{303}, 2 (2003)}.}
	%
	\bibitem []{Nayak2008}%
	{C.~Nayak, S.~H. Simon, A.~Stern, M.~Freedman, and S.~Das Sarma, \href{\doibase
			10.1103/RevModPhys.80.1083}{Rev. Mod. Phys. \textbf{80}, 1083 (2008).}} 
		%
	\bibitem []{Wilczek2009}%
	{F.~Wilczek, \href{\doibase 10.1038/nphys1380} {Nat. Phys. \textbf{5}, 61 (2009).}} 
	%
	\bibitem []{Stern}%
	{A.~Stern, \href{\doibase 10.1038/nature08915}{Nature \textbf{464}, 187 (2010).}} 
	%
	\bibitem []{Schnyder2008}%
	{A.~P. Schnyder, S.~Ryu, A.~Furusaki, and A.~W.~W. Ludwig, \href
		{\doibase 10.1103/PhysRevB.78.195125} {Phys. Rev. B \textbf{78}, 195125 (2008).}} 
	%
	\bibitem []{Sato2009}%
	{M.~Sato and S.~Fujimoto, \href{\doibase 10.1103/PhysRevB.79.094504}{Phys. Rev. B \textbf{79}, 094504 (2009).}}
	%
	\bibitem []{Qi2011}%
	{X.-L. Qi and S.-C. Zhang, \href {\doibase 10.1103/RevModPhys.83.1057} {Rev. Mod. Phys. \textbf{83}, 1057 (2011).}}
	 %
	\bibitem []{Tanaka2012}%
	{Y.~Tanaka, M.~Sato, and N.~Nagaosa, \href{\doibase 10.1143/JPSJ.81.011013} {J. Phys. Soc. Japan\ \textbf{81}, 011013 (2012).}} 
	%
	\bibitem []{Vijay2015}%
	{S.~Vijay, T.~H. Hsieh, and L.~Fu, \href{\doibase 10.1103/PhysRevX.5.041038}{Phys. Rev. X \textbf{5}, 041038 (2015).}} 
	%
	\bibitem []{Vijay2016}%
	{S.~Vijay and L.~Fu, \href{\doibase 10.1103/PhysRevB.94.235446}{Phys. Rev. B \textbf{94}, 235446 (2016).}} 
	%
	\bibitem []{Cheng2012}%
	{M.~Cheng, R.~M. Lutchyn, and S.~Das Sarma, \href {\doibase 10.1103/PhysRevB.85.165124} {Phys. Rev. B \textbf{85}, 165124 (2012).}} 
	%
	\bibitem []{Bonderson2013}%
	{P.~Bonderson and C.~Nayak, \href{\doibase 10.1103/PhysRevB.87.195451}{Phys. Rev. B \textbf{87}, 195451 (2013).}} 
	%
	\bibitem []{Sarma2015}%
	{S.~D. Sarma, M.~Freedman, and C.~Nayak, \href{http://dx.doi.org/10.1038/npjqi.2015.1} {npj Quantum Inf. \textbf{1}, 15001 (2015).}}
	%
	\bibitem []{Nadj-Perge2014}%
	{S.~Nadj-Perge, I.~K. Drozdov, J.~Li, H.~Chen, S.~Jeon, J.~Seo, A.~H. MacDonald, B.~A. Bernevig, and A.~Yazdani, \href {\doibase 10.1126/science.1259327}
		{Science \textbf{346}, 602 (2014).}} 
	%
	\bibitem []{Pawlak2016}%
	{R.~Pawlak, M.~Kisiel, J.~ Klinovaja, T.~ Meier,  S.~Kawai, T.~ Glatzel,
				D.~Loss, \ and\   
				E.~Meyer, \href {\doibase 10.1038/npjqi.2016.35}
		{npj Quantum Inf. \textbf{2},   16035 (2016).}} 
	%
	\bibitem []{Ruby2015}%
	{M.~Ruby,  F.~Pientka, Y.~Peng,  F.~von Oppen, B.~W. Heinrich, and  K.~J. Franke,  \href {\doibase
			10.1103/PhysRevLett.984808} { Phys. Rev. Lett. \textbf{115}, 197204 (2015).}} 
		%
	\bibitem []{Alicea2010}%
	{J.~Alicea, \href{\doibase 10.1103/PhysRevB.81.125318} {Phys. Rev. B \textbf{81}, 125318 (2010).}}
	%
	\bibitem []{Lutchyn2010}%
	{R.~M. Lutchyn, J.~D. Sau, and S.~Das Sarma, \href{\doibase 10.1103/PhysRevLett.105.077001}{Phys. Rev. Lett. \textbf{105}, 077001 (2010).}} 
%
	\bibitem []{Oreg2010}%
	{Y.~Oreg, G.~Refael, and F.~von Oppen, \href
		{\doibase 10.1103/PhysRevLett.105.177002}{Phys. Rev. Lett. \textbf{105} 177002 (2010).}}
 %
	\bibitem []{Mourik2012}%
	{V.~Mourik, K.~Zuo, S.~M. Frolov, S.~R. Plissard, E.~P. A.~M. Bakkers, and L.~P. Kouwenhoven, \href
		{\doibase 10.1126/science.1222360}{Science \textbf{336}, 1003 (2012).}}
	 %
	\bibitem []{Das2012}%
	{A.~Das, Y.~Ronen, Y.~Most, Y.~Oreg, M.~Heiblum, and H.~Shtrikman, \href{\doibase 10.1038/nphys2479}
		{Nat. Phys. \textbf{8}, 887 (2012).}}
	%
	\bibitem []{Deng2012}%
	{M.~T. Deng, C.~L. Yu, G.~Y.  Huang, M.~Larsson, P.~Caroff, and H.~Q.  Xu, \href{\doibase
			10.1021/nl303758w} { Nano Lett. \textbf{12}, 6414 (2012).}} 
		%
	\bibitem []{Sticlet2012}%
	{D.~Sticlet, C.~Bena, and P.~Simon, \href{\doibase 10.1103/PhysRevLett.108.096802} {Phys. Rev. Lett. \textbf{108}, 096802 (2012). }} 
%
	\bibitem []{Rokhinson2012}%
	{L.~P. Rokhinson, X.~Liu, and J.~K. Furdyna, \href {\doibase 10.1038/nphys2429} {Nat. Phys. \textbf{8}, 795 (2012). }} 
%
	\bibitem []{Klinovaja2012}%
	{J.~Klinovaja and  D.~Loss, \href {\doibase 10.1103/PhysRevB.86.085408} {  
			Phys. Rev. B \textbf{86},
			085408 (2012).}} 
		%
	\bibitem []{Chevallier2012}%
	{D.~Chevallier, D.~Sticlet,
			P.~ Simon, \ and\ 
			C.~Bena, \href {\doibase
			10.1103/PhysRevB.85.235307} {Phys.
					Rev. B \textbf{85},  235307
			(2012).}} 
		%
	\bibitem []{San-Jose2012}%
	{P.~San-Jose,   E.~ Prada, 
			and R.~Aguado, \href
		{\doibase 10.1103/PhysRevLett.108.257001} {Phys. Rev. Lett. \textbf{108}, 257001 (2012). }}
	 %
	\bibitem []{Dominguez2012}%
	{F.~Dom{\'i}nguez, F.~Hassler, and G.~Platero, \href {\doibase 10.1103/PhysRevB.86.140503} {Phys. Rev. B \textbf{86}, 140503 (2012).}}
	%
	\bibitem []{Terhal2012}%
	{B.~M. Terhal, F.~Hassler, and D.~P. DiVincenzo, \href{\doibase 10.1103/PhysRevLett.108.260504} {  
			Phys. Rev. Lett. \textbf{108}, 260504 (2012).}} 
		%
	\bibitem []{Klinovaja2012a}%
	{J.~Klinovaja, P.~Stano, and D.~Loss, \href
		{\doibase 10.1103/PhysRevLett.109.236801} {Phys. Rev. Lett. \textbf{109} 236801 (2012).}} 
	%
	\bibitem []{Prada2012}%
	{
		{  { {E.}~
				{Prada}},   { {P.}~ {San-Jose}}, \
			and\   { {R.}~ {Aguado}},\ }\href
		{\doibase 10.1103/PhysRevB.86.180503} {   {
				{Phys. Rev. B}\ }\textbf {  {86}},\ 
			{180503} (2012).}} 
		%
	\bibitem []{Churchill2013}%
	{H.~O.~H. Churchill, V.~Fatemi, K.~Grove-Rasmussen, M.~T. Deng, P.~Caroff, H.~Q. Xu, and C.~M. Marcus, \href {\doibase
			10.1103/PhysRevB.87.241401} {Phys. Rev. B \textbf{87}, 241401
			(2013).}} 
		%
	\bibitem []{DeGottardi2013}%
	{W.~DeGottardi, M.~Thakurathi, S.~Vishveshwara, and D.~Sen, \href {\doibase 10.1103/PhysRevB.88.165111} {  
			Phys. Rev. B \textbf{88}, 165111 (2013).}} %
	\bibitem []{Thakurathi2013}%
	{ M.~Thakurathi, A.~A. Patel, D.~Sen, and A.~Dutta, \href
		{\doibase 10.1103/PhysRevB.88.155133}{Phys. Rev. B \textbf{88}, 155133 (2013).}} 
	%
	\bibitem []{Maier2014}%
	{F.~Maier, J.~Klinovaja, and D.~Loss, \href
		{\doibase 10.1103/PhysRevB.90.195421} {Phys. Rev. B \textbf{90}, 195421 (2014).}} 
	%
	\bibitem []{Escribano2017}%
	{S.~D. Escribano, A.~L. Yeyati, and E.~Prada, \href{\doibase 10.3762/bjnano.9.203}{Beilstein J. Nanotechnol. \textbf{9}, 2171 (2018).}}
	%
	\bibitem []{Prada2017}%
	{E.~Prada, R.~Aguado, and P.~San-Jose, \href
		{\doibase 10.1103/PhysRevB.96.085418} {Phys. Rev. B \textbf{96}, 085418 (2017).}}
	 %
	\bibitem []{Ptok2017}%
	{A.~Ptok, A.~Kobia{\l}ka, and T.~Doma{\'{n}}ski, \href{\doibase 10.1103/PhysRevB.96.195430} {Phys. Rev. B \textbf{96}, 195430 (2017).}}
	 %
	\bibitem []{Kobialka2018}%
	{A.~Kobia{\l}ka and A.~Ptok, \href{\doibase 10.1088/1361-648X/ab03bf/meta}{J. Phys. Condens. Matter \textbf{31}, 18 (2019).}} %
	\bibitem []{DeMoor2018}%
	{M.~W.~A. de~Moor, J.~D.~S. Bommer, D.~Xu, G.~W. Winkler, A.~E. Antipov, A.~Bargerbos, G.~Wang, N.~van Loo, R.~L.~M. Op het Veld, S.~Gazibegovic, D.~Car, J.~A. Logan, M.~Pendharkar, J.~S. Lee, E.~P.~A.~M. Bakkers, C.~J. Palmstr{\o}m, R.~M. Lutchyn, L.~P. Kouwenhoven, and H.~Zhang, \href{\doibase
			10.1088/1367-2630/aae61d}{New J.
					Phys. \textbf{20}, 103049 (2018).}}
				 %
	\bibitem []{Budich2012}%
	{J.~C. Budich, S.~Walter, and B.~Trauzettel, \href{\doibase 10.1103/PhysRevB.85.121405} {Phys. Rev. B \textbf{85}, 121405 (2012).}} 
	%
	\bibitem[]{Scheurer2013}{M. S. Scheurer and A. Shnirman, \href{\doibase 10.1103/PhysRevB.88.064515}{Phys. Rev. B \textbf{88}, 064515 (2013).}}
	%
	\bibitem[]{Sekania2017}
	{M.~Sekania, S.~Plugge, M.~Greiter, R.~Thomale, and P.~Schmitteckert, \href{\doibase 10.1103/PhysRevB.96.094307}{Phys. Rev. B \textbf{96}, 094307 (2017).} }
	\bibitem []{Goldstein2011}%
	{G.~Goldstein and C.~Chamon, \href{\doibase 10.1103/PhysRevB.84.205109}{Phys. Rev. B \textbf{84}, 205109 (2011).}}
	%
	\bibitem[]{Pedrochi2015}{F.~L.~Pedrocchi and D.~P.~DiVincenzo, \href{\doibase 10.1103/PhysRevLett.115.120402}{Phys. Rev. Lett. \textbf{115}, 120402 (2015).}}
	%
	\bibitem[]{Pedrochi2015b}
	{F.~L. Pedrocchi, N.~E. Bonesteel, and D.~P. DiVincenzo, \href{https://journals.aps.org/prb/abstract/10.1103/PhysRevB.92.115441}{Phys. Rev. B \textbf{92}, 115441 (2015).}}
	%
	\bibitem []{Rainis2012}%
	{D.~Rainis and D.~Loss, \href{\doibase 10.1103/PhysRevB.85.174533}{Phys. Rev. B \textbf{85}, 174533 (2012).}}
	 %
	\bibitem []{Knapp2018}%
	{C.~Knapp, T.~Karzig, R.~M. Lutchyn, and C.~Nayak, \href
		{\doibase 10.1103/PhysRevB.97.125404}{Phys. Rev. B \textbf{97}, 125404 (2018).}} 
	%
	\bibitem []{Schmidt2012}%
	{M.~J.~Schmidt, D.~Rainis, and D.~Loss, \href
		{\doibase 10.1103/PhysRevB.86.085414}{Phys. Rev. B \textbf{86}, 085414 (2012).}}
	 %
	\bibitem []{Lai2018}%
	{H.-L. Lai, P.-Y. Yang, Y.-W. Huang, and W.-M. Zhang, \href{\doibase 10.1103/PhysRevB.97.054508}{Phys. Rev. B \textbf{97}, 054508 (2018).}} 
	%
	\bibitem []{Aseev2018}%
	{P.~P. Aseev, J.~Klinovaja, and D.~Loss, \href
		{\doibase 10.1103/PhysRevB.98.155414} {Phys. Rev. B \textbf{98}, 155414 (2018).}} %
	\bibitem []{Szumniak2017}{
		P. Szumniak, D. Chevallier, D. Loss, and J. Klinovaja, \href{\doibase 10.1103/PhysRevB.96.041401}{Phys. Rev. B \textbf{96}, 041401(R) (2017).}
	}
	
	\bibitem []{Cleland2002}%
	{A.~N. Cleland, \href{https://books.google.co.jp/books?id=lM1Uq53RU-EC} {\emph{Foundations of Nanomechanics: From Solid-State Theory to
						Device Applications}}, Advanced Texts in Physics ({Springer Berlin Heidelberg},  {2002}).} %
	\bibitem []{Trif2008}
	%
	{M.~Trif, V.~N. Golovach,
			and   D.~Loss, \href
		{\doibase 10.1103/PhysRevB.77.045434} { 
				Phys. Rev. B \textbf {77}, 45434 (2008).}} 
			%
	\bibitem []{Kloeffel2014}%
	{C.~Kloeffel, M.~Trif, and D.~Loss, \href
		{\doibase 10.1103/PhysRevB.90.115419}{Phys. Rev. B \textbf{90}, 115419 (2014).}} 
	%
	\bibitem []{Gangadharaiah}%
	{S.~Gangadharaiah, B~Braunecker, P.~Simon, and D.~Loss, \href{\doibase 10.1103/PhysRevLett.107.036801}{Phys. Rev. Lett. \textbf{107}, 036801 (2011).}} 
	%
	\bibitem{AGD}{A. A. Abrikosov, L. P. Gorkov, and I. E. Dzialoshinskii,
		\textit{Quantum field theoretical methods in statistical physics},
		Vol. 4, Pergamon (1965).}
	%
	\bibitem []{Mahan2000}%
	{
		{  { {G.~D.}\ 
				{Mahan}},\ } {\emph {  {{Many-particle
						physics}}}},\  {edition} {3rd}\ ed.\ ({Kluwer
			Academic/Plenum Publishers},\   {2000}).}
		 %
	\bibitem []{Keldysh1965diagram}%
	{L.~V. Keldysh, \href
		{http://www.jetp.ac.ru/cgi-bin/dn/e{\_}020{\_}04{\_}1018.pdf} {
			Sov. Phys. JETP \textbf{20}, 1018 (1965).}}
	\bibitem{Kamenev09}{ A. Kamenev and A. Levchenko, \href{https://www.tandfonline.com/doi/abs/10.1080/00018730902850504}{Advances in Physics {\bf 58}, 197 (2009).}
}
	%
	\bibitem []{Klemens1966}%
	{P.~G. Klemens, \href{\doibase 10.1103/PhysRev.148.845}{Phys. Rev. \textbf{148}, 845 (1966).}}
	 %
	\bibitem []{Balkanski1983}%
	{M.~Balkanski, R.~F. Wallis, and E.~Haro, \href {\doibase 10.1103/PhysRevB.28.1928} {  
		Phys. Rev. B \textbf{28}, 1928 (1983).}} 
	%
	\bibitem []{Menendez1984}%
	{J.~Men{\'{e}}ndez and M.~Cardona, \href{\doibase 10.1103/PhysRevB.29.2051} {  
			{Phys. Rev. B} \textbf{29}, 2051 (1984).}} %
	
	\bibitem []{Rainis2013}		
{D. Rainis, L. Trifunovic, J. Klinovaja, and D. Loss, \href {\doibase
		10.1103/PhysRevB.87.024515}{Phys. Rev. B {\bf 87}, 024515 (2013).}}
			
\end{thebibliography}
\end{document}